## Generation of planar tensegrity structures through cellular multiplication

Omar ALOUI[a], David ORDEN[b], Landolf RHODE-BARBARIGOS*

[a] Department of Civil, Architectural & Environmental Engineering, University of Miami, 1251 Memorial Drive, Coral Gables, FL 33146-0630, USA, Email: omar.aloui@miami.edu

[b] Departamento de Física y Matemáticas, Universidad de Alcalá. Ctra. Madrid-Barcelona, Km. 33,600, 28805, Alcalá de Henares, Spain, Email: david.orden@uah.es

* Department of Civil, Architectural & Environmental Engineering, University of Miami, 1251 Memorial Drive, Coral Gables, FL 33146-0630, USA, Email: landolfrb@miami.edu

### Abstract

Tensegrity structures are frameworks in a stable self-equilibrated prestress state that have been applied in various fields in science and engineering. Research into tensegrity structures has resulted in reliable techniques for their form finding and analysis. However, most techniques address topology and form separately. This paper presents a bio-inspired approach for the combined topology identification and form finding of planar tensegrity structures. Tensegrity structures are generated using tensegrity cells (elementary stable self-stressed units that have been proven to compose any tensegrity structure) according to two multiplication mechanisms: cellular adhesion and fusion. Changes in the dimension of the self-stress space of the structure are found to depend on the number of adhesion and fusion steps conducted as well as on the interaction among the cells composing the system. A methodology for defining a basis of the self-stress space is also provided. Through the definition of the equilibrium shape, the number of nodes and members as well as the number of self-stress states, the cellular multiplication method can integrate design considerations, providing great flexibility and control over the tensegrity structure designed and opening the door to the development of a whole new realm of planar tensegrity systems with controllable characteristics.

## 1. Introduction

### 1.1. Definitions and applications

Tensegrity describes a system with components in a prestressed self-equilibrated state. The term was first introduced by Buckminster Fuller in 1962 to describe Kenneth Snelson's sculptures [1,2]. Tensegrity systems generated lots of interest from mathematicians [3,4], architects [5,6], material [7], structural [8,9], aerospace [10,11], robotics [12,13] and biomechanical engineers [14,15]. The broad spectrum of tensegrity applications has also been reflected by the diverse set of definitions employed to describe the concept. Definitions vary from field to field and often from one researcher to another. The most rigorous definition in engineering and architecture was provided by Motro (2005) [2]. He defined tensegrity as "*a system in stable self-equilibrated state comprising a discontinuous set of compressed components inside a continuum of tensioned components*". In mathematics, tensegrity has been defined as a self-stressed framework [3]. A framework is a realization of an abstract graph $G(V,E)$ described by a set of vertices $V=\{v_1,v_2,\ldots,v_n\}$ and the pairs of vertices $E$ in a d-dimensional space through a finite configuration $P=\{p_1,p_2,\ldots p_n\}$ where $P$ is a collection of coordinates $p_i$ assigned to each vertex $v_i$. In a planar framework, $p_i(p_{ix},p_{iy})$ represents the $x$ and $y$ coordinates of a node in the structure. In this paper, the mathematical definition is considered as it represents a generalization that encompasses most existing definitions including the one provided by Motro [2].

### 1.2. Form finding and topology identification of tensegrity structures

The first step in the design of tensegrity systems is typically form finding. Form finding describes the forward process of finding a stable equilibrium configuration for a structure under a specific set of loading and boundary conditions starting from an arbitrary initial geometry [16]. For a tensegrity system, form finding reflects the identification of a stable self-equilibrium configuration under prestress with only rigid body motions constrained. Form-finding methods, such as force density [17,18] and dynamic relaxation [19,20], have been employed for the form finding and analysis of tensegrity structures investigating directly or indirectly the relation between topology, geometry and structural behavior in the resulting structures. However, force density and dynamic relaxation require that topology and connectivity of the system to be predefined. In engineering applications, member typology (tension/compression or tension solely) is also often required as an input. Consequently, topology definition and form finding in tensegrity systems typically reflect two independent problems often addressed in this order resulting in systems that may not be optimal or appropriate for the given problem. Zhang et al. [21] and Lee et al. [22] tried to address this issue by combining the force-density method with a genetic algorithm so that the form-finding process requires minimum knowledge about the connectivity of the structure and the member typology is not a required input parameter. Nevertheless, due to the combinatorial nature of the connectivity problem in tensegrity structures, brute-force approaches tackling this problem are computationally expensive.

Self-stress (or force mode) describes a set of forces that induces a state of self-equilibrium in a tensegrity structure. It is a key feature of tensegrity structures as it defines their form, properties and behavior. The importance of the self-stress resides in the information it holds regarding the properties of the structure. For example, modifying self-stress according to an equilibrium manifold (a continuous set of solutions of the self-stress problem) can allow a tensegrity structure to remain stable throughout shape transformations [23]. Changes in the shape of tensegrity structures can thus be controlled by element-length modifications obtained through the integration of active elements [8-10] or smart materials [27-29]. Moreover, since the self-stress guarantees stability in tensegrity structures, the existence of multiple self-stress states increases the chances of survival after a member removal. Consequently, damage tolerance is another property closely related to self-stress with the number of states corelating with the degree of indeterminacy in a tensegrity structure. Therefore, the description of the self-stress space as well as the identification of a self-stress state that satisfies member typology (conform state) are important steps in the design of tensegrity systems. The space for the possible combinations that describe conform self-stress solutions can be found using linear programming techniques like the vertex-enumeration problem [27,28] and elements grouping methods [29]. However, the existence of such solutions depends on the topology and configuration (geometry) of the system, which in the aforementioned methods are required as input limiting the spectrum of structures most often to simple regular structures. Ehara and Kanno [30,31,32] tried to design irregular tensegrity structures by using mixed integer linear programming (MILP) which showed a promising potential in finding class 1 tensegrity structures but did not address class k tensegrity structures (tensegrity systems with a maximum of k interconnected compressive members [33]).

Many researchers have tried to dissociate the link between topology and configuration (geometry) in the self-stress problem. Rieffel et al. [34] addressed the self-stress problem of irregular tensegrity configurations through grammar-based representations graphs. Xu et al. [35] used the ground structure method and proposed a refinement for the MILP method into mixed integer quadratic programming (MIQP). Xu et al. [36] upgraded the method and combined force density method with mixed nonlinear integer programming to find feasible topologies and geometries for irregular tensegrity structures. Lee and Lee [22] combined force density method with a genetic algorithm to define the topology of tensegrity structures requiring nodal coordinates as input. Li et al. [37] adopted a similar approach by growing tensegrity structures using one-bar units. Although these methods solve the self-stress problem, they are limited to specific classes of tensegrity systems. In addition, they do not provide direct control on the nature nor the number of self-stress states.

The characterization of tensegrity-system topologies and the relation between connectivity, configuration and the redundancy of the structure has been studied in graph theory and rigidity theory. Laman [38] first characterized minimally generically rigid frameworks in the plane and introduced the class of graphs known as Laman graphs, where the removal of one element introduces a mechanism in the structure. However, tensegrity systems are self-stressed frameworks with their rigidity being also influenced by member typology. A characterization of minimally generically rigid tensegrity structures was proposed by Recski [39]: A simple graph $G$ with $n$ vertices and $2n$-$2$ edges is a generically rigid tensegrity in the plane if and only if $|E'| \leq 2|V'| - 3$ holds for every proper subgraph $G'(V',E')$ of G with at least two vertices. De Guzmán and Orden [4] proposed a decomposition algorithm for tensegrity structures into elementary stable units, which Fernández and Orden [40] further developed to characterize the relation between these units and the changes in the self-stress space.

This paper builds upon those results and presents a novel bio-inspired approach for the generation of planar tensegrity structures. The analogy between biological cells and tensegrity structures has already been used in biomechanics, with tensegrity structures being used as structural models for the investigation of the biomechanical behavior of living cells [41-42]. However, in this paper, it is the multiplication mechanisms of biological cells that are employed to explain the topological and geometrical characteristics of tensegrity structures. The method is thus inspired by cellular multiplication mechanisms and does not require prior knowledge on the topology of the tensegrity structure, while providing control on key parameters such as the number of nodes, members, self-stress states and the equilibrium shape (form) of the structure. The method resolves the self-stress states in a planar tensegrity structure through its decomposition in minimally rigid tensegrity systems, providing an intuitive approach for understanding the stability and the redundancy of tensegrity structures. The remainder of the paper is organized as follows: in Section 2, the self-stress problem is presented along with foundational theorems and propositions. The cellular multiplication method and its mechanisms are described in Section 3. In Sections 4 and 5, the update of the self-stress space during the cellular multiplication process and the interaction among tensegrity cells are analyzed. Examples of tensegrity structures generated with the cellular multiplication method are presented in Section 6. Further discussion and conclusions are found in Sections 7 and 8.

## 2. Self-stress in tensegrity structures

### 2.1. Self-stress definition and importance

Most definitions on tensegrity systems agree upon the fact that they are frameworks in a state of stable self-equilibrium defined by a set of internal forces induced by the topological and the geometrical interactions of the structural members without considering external forces or reactions [2,43]. Depending on the system's topology (when the underlying graph of the structure is generically rigid) and its geometry (if the configuration of the structure contains degenerate positions), a tensegrity structure can have multiple self-stress states which define a subspace of all the possible force assignments in its members. To understand the self-stress problem, the equilibrium problem of the structure is addressed first.

Let $E$ be the set of members of the structure and $|E|$ the total number of members. Let $V$ be the set of nodes of the structure and $|V|$ the total number of nodes. Let $\overline{x_i}$ be the vector of Cartesian coordinates of a node $i$. Considering $\overline{p_i}$ as the vector of the external force at node $i$ and $w_{ij}$ as a scalar representing the force over the length of the element (force density) linking node $i$ to $j$, the equilibrium at node $i$ of the structure is given by [44]:

$$\sum_{\substack{j \\ (i,j)\in E}} (\overline{x}_i - \overline{x}_j) w_{ij} = \overline{p}_i \quad (1)$$

The equilibrium at every node results in a system of $d|V|$ equations that can be described algebraically as:

$$Aw = p \quad (2)$$

where w is a vector of $|E|$ force components, $p$ is a vector of $d|V|$ components of nodal forces and $A$ is the equilibrium matrix. Self-stress can then be defined as the set of forces that induces a state of self-equilibrium in the structure without considering external loads or supports:

$$Aw = 0 \quad (3)$$

The self-stress space is characterized algebraically by the null space of the equilibrium matrix $A$:

$$W = nullspace(A) \quad (4)$$

where $W$ is a basis of the space. This characterization provides an algebraic method for the calculation of self-stress. However, the use of this method requires the definition of the equilibrium matrix and thus the existence of a stable configuration. An alternative combinatorial characterization of the self-stress space can be found in rigidity theory where the self-stress state in a framework *(E,V,P)* corresponds to an assignment of scalars to each pair of vertices *(i,j)* such that for each vertex $v_i$, the scaled sum of incident vectors is 0 (nodal equilibrium condition).

## 2.2. Foundational theorem and propositions

The generation method proposed in this paper is based on a series of statements, theorems and propositions, adapted from rigidity theory and graph theory. The first statement is a theorem developed by de Guzmán and Orden [4] focusing on the decomposition of tensegrity structures into elementary units:

> **Theorem.** Let *T(P)* be the tensegrity structure defined by the framework *(E,V,P)* where *G=(E,V)* is the abstract graph on the set of vertices *V* and the set of edges *E*, and *P* is a configuration of points in d-space in general position (with no $d+1$ points lying on the same hyperplane). The tensegrity structure *T(P)* is then a finite sum of elementary units defined by the complete graph on $d+2$ points, denoted $K_{d+2}$.

The decomposition can also be employed to check if a graph corresponds to a tensegrity structure and the solution is not unique, as multiple decomposition paths may exist. Cellular multiplication of tensegrity structures employs the elementary units described in the theorem as constitutive elements to compose complex tensegrity structures while also constructing a base describing their self-stress space.

The second statement characterizes combinatorically the dimension of the self-stress space. In rigidity theory, the dimension of the self-stress space $|W|$ is related to the number of degrees of freedom (infinitesimal mechanisms) of the framework *G(P)* denoted *df* by the Proposition I below proposed by Graver et al. [45]:

> **Proposition I.** Let *G(P)* be a framework in general position *P* in dimension d with *G=(E,V)* the underlying abstract graph of the framework and $|W|$ the dimension of its self-stress space. The number of degrees of freedom *df* of the framework *G(P)* is given by:

$$df = \begin{cases} |W| - (\frac{d(d+1)}{2} + |E| - d|V|) & \text{if } |V| \geq d \\ \frac{|V|(|V|-1)}{2} - |E| & \text{if } |V| \leq d \end{cases}$$

Proposition I expresses thus the generalized Maxwell's counting rule for the determinacy of frameworks proposed by Calladine [46]:

$$s - m = b - d\,j + \frac{d(d+1)}{2} \quad (5)$$

where $s$ is the number of self-stress states given by $|W|$ in the proposition, $m$ is the number of infinitesimal mechanisms also known as the degrees of freedom $df$, $b$ is the number of bars of the framework, $j$ its number of joints and $d$ is the dimension of the workspace. In Proposition I, the second part of the difference corresponds to the Laman bound $B$:

$$B = \frac{d(d+1)}{2} + |E| - d|V| \quad (5)$$

The Laman bound of a graph $G$ represents the dimension of the self-stress space of the graph in a generic position. Fernández and Orden [40] employed Proposition I along with the decomposition algorithm described in de Guzmán and Orden [4] to describe recursively the dimension of the self-stress space in tensegrity structures:

> **Proposition II.** Let $T_i$ and $T_{i+1}$ be two successive intermediate tensegrity sub-structures in the combinatorial decomposition of a tensegrity structure in dimension d, with every vertex in $T_i$ having degree at least $d$. The Laman bound $B_{i+1}$ of the sub-structure $T_{i+1}$ is given by:
>
> $$B_{i+1} = B_i + e_i - 1$$
>
> where $B_i$ is the Laman bound of the sub-structure $T_i$, and $e_i$ is the number of edges required to pass from the sub-structure $T_i$ to sub-structure $T_{i+1}$.

If the underlying graph of the tensegrity structure is generically rigid, then the dimension of the self-stress space $|W|$ is exactly the Laman bound $B$ of the graph. This result allows the combinatorial calculation of $|W|$ using the decomposition of the structure to the elementary units denoted as tensegrity cells in this paper.

### 2.3. Planar tensegrity cells

Planar tensegrity cells are identified as the complete graphs $K_4$ on four nodes and are illustrated in Figure 1. A graph is called complete when all pairs of vertices are connected by an edge. Although topologically the two cells are the same, the embedding of the abstract graph $K_4$ in the plane results in two different structures according to member typology. Elements in a Type I cell can be classified into two groups of same type (tension or compression): a group of 4 elements and a group of 2 elements. On the other hand, Type II cell has two groups of 3 elements. In Figure 1, element groups are distinguished using solid and dashed lines. It should be noted that member typology is not required to be assigned at this stage, as groups can take compression or tension resulting in four total different structures with the consideration of the duals.

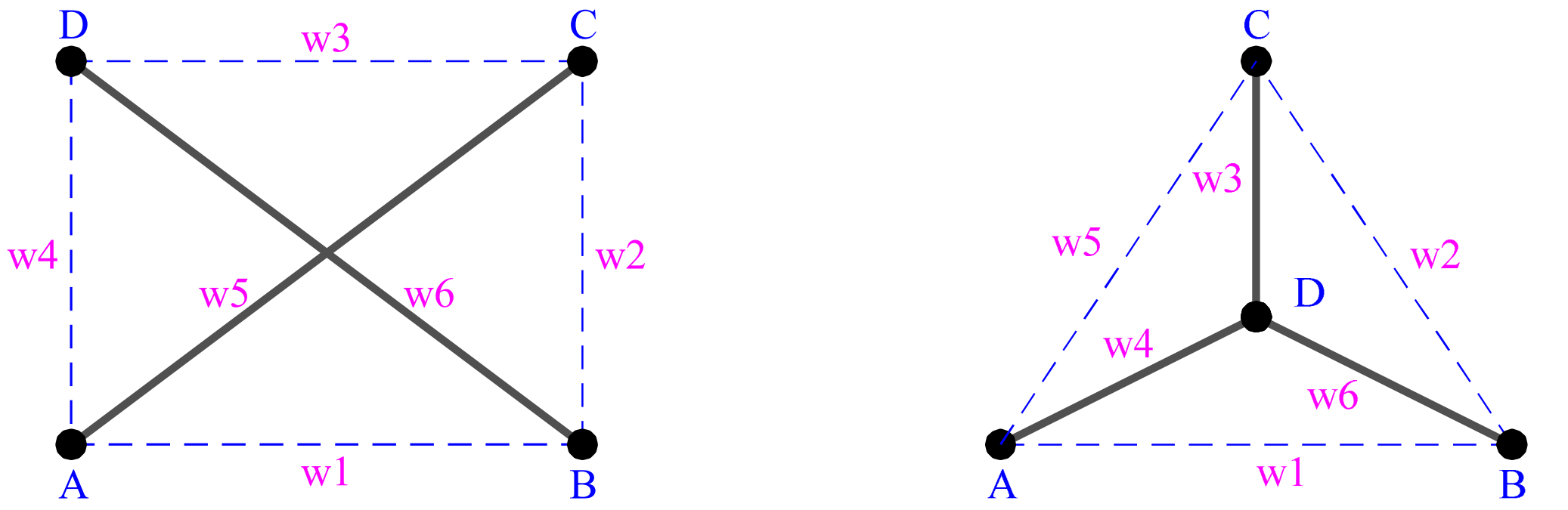

Type I Type II

Figure 1: Illustration of planar tensegrity cells. Element groups are distinguished using solid and dashed lines.

Tensegrity cells have one stable self-stress state and they are infinitesimally rigid (no infinitesimal mechanisms). The analytical solution of their self-stress is provided below. Let $A(a_1,a_2)$, $B(b_1,b_2)$, $C(c_1,c_2)$ and $D(d_1,d_2)$ be the configuration of the tensegrity cell in the plane and $O$ the origin. Let $w_1,w_2,\ldots,w_6$ be the force densities (force in the element divided by the length of the element) assigned to each member (Figure 1). Writing the nodal equilibrium for both Type I and Type II cells gives:

$$w_1\overrightarrow{AB}+w_4\overrightarrow{AD}+w_5\overrightarrow{AC}=\vec{0} \tag{6}$$

$$w_2\overrightarrow{CB}+w_3\overrightarrow{CD}+w_5\overrightarrow{CA}=\vec{0} \tag{7}$$

$$w_1\overrightarrow{BA}+w_2\overrightarrow{BC}+w_6\overrightarrow{BD}=\vec{0} \tag{8}$$

$$w_4\overrightarrow{DA}+w_3\overrightarrow{DC}+w_6\overrightarrow{DB}=\vec{0} \tag{9}$$

The system has exactly one non-trivial solution. Without loss of generality, assume that $w_1$ is known. Applying the 2D cross product of Equation (6) and vector $\overrightarrow{AC}$ gives:

$$\begin{aligned}
&(w_1\overrightarrow{AB}+w_4\overrightarrow{AD}+w_5\overrightarrow{AC})\times\overrightarrow{AC}\Leftrightarrow w_4\overrightarrow{AD}\times\overrightarrow{AC}=-w_1\overrightarrow{AB}\times\overrightarrow{AC}\\
&\Leftrightarrow w_4(\overrightarrow{AO}+\overrightarrow{OD})\times(\overrightarrow{AO}+\overrightarrow{OC})=-w_1(\overrightarrow{AO}+\overrightarrow{OB})\times(\overrightarrow{AO}+\overrightarrow{OC})\\
&\Leftrightarrow w_4(\overrightarrow{OA}\times\overrightarrow{OC}-\overrightarrow{OA}\times\overrightarrow{OD}+\overrightarrow{OC}\times\overrightarrow{OD})=w_1(\overrightarrow{OA}\times\overrightarrow{OB}-\overrightarrow{OA}\times\overrightarrow{OC}+\overrightarrow{OB}\times\overrightarrow{OC})\\
&\Leftrightarrow w_4\left(\begin{vmatrix}a_1 & a_2\\ c_1 & c_2\end{vmatrix}-\begin{vmatrix}a_1 & a_2\\ d_1 & d_2\end{vmatrix}+\begin{vmatrix}c_1 & c_2\\ d_1 & d_2\end{vmatrix}\right)=w_1\left(\begin{vmatrix}a_1 & a_2\\ b_1 & b_2\end{vmatrix}-\begin{vmatrix}a_1 & a_2\\ c_1 & c_2\end{vmatrix}+\begin{vmatrix}b_1 & b_2\\ c_1 & c_2\end{vmatrix}\right)
\end{aligned} \tag{10}$$

Introducing the function $f$ with $\frac{1}{2}|f|$ corresponding to the affine area of the triangle $A,B,C$:

$$\mathrm{f(A,B,C)}=\begin{vmatrix}a_1 & a_2\\ b_1 & b_2\end{vmatrix}-\begin{vmatrix}a_1 & a_2\\ c_1 & c_2\end{vmatrix}+\begin{vmatrix}b_1 & b_2\\ c_1 & c_2\end{vmatrix}=\begin{vmatrix}1 & a_1 & a_2\\ 1 & b_1 & b_2\\ 1 & c_1 & c_2\end{vmatrix} \tag{11}$$

$w_4$ can be written as:

$$w_4 = w_1 \frac{\begin{vmatrix} 1 & a_1 & a_2 \\ 1 & b_1 & b_2 \\ 1 & c_1 & c_2 \end{vmatrix}}{\begin{vmatrix} 1 & a_1 & a_2 \\ 1 & c_1 & c_2 \\ 1 & d_1 & d_2 \end{vmatrix}} = w_1 \frac{\mathrm{f(A,B,C)}}{\mathrm{f(A,C,D)}} \tag{12}$$

Reflecting the area of triangle *A,B,C*, the function *f(A,B,C)* is null if and only if *A,B,C* lie on the same line, which is useful for the detection of the configurations of the cell that are not in general position. This property of function f derives also from its definition as the cross product $-\overrightarrow{AB} \times \overrightarrow{AC}$. Repeating the process for Equations (7) to (9) allows to find the other self-stress components, and the self-stress state $w$ can be written as:

$$w = \begin{bmatrix} w_1 \\ w_2 \\ w_3 \\ w_4 \\ w_5 \\ w_6 \end{bmatrix} = w_1 \begin{bmatrix} 1 \\ \dfrac{\mathrm{f(A,B,D)}}{\mathrm{f(B,C,D)}} \\ \left( \dfrac{\mathrm{f(A,B,D)}}{\mathrm{f(A,C,D)}} \cdot \dfrac{\mathrm{f(A,B,C)}}{\mathrm{f(B,}C\mathrm{,D)}} \right) \\ \dfrac{\mathrm{f(A,B,C)}}{\mathrm{f(A,C,D)}} \\ -\dfrac{\mathrm{f(A,B,D)}}{\mathrm{f(A,C,D)}} \\ -\dfrac{\mathrm{f(A,B,C)}}{\mathrm{f(B,C,D)}} \end{bmatrix} \tag{13}$$

## 3. Multiplication of planar tensegrity cells

### 3.1. Cellular multiplication analogy

The multiplication of tensegrity cells is inspired by the natural multiplication process of cellular organisms. In nature, adhesion occurs when cells attach to their neighboring cells to form complex multicellular organisms. This process results in two similar yet distinct entities that can survive on their own (they are stable and function separately). If stable cells are connected together and interact through their membrane, then this process is called cellular adhesion. Cellular fusion is another process in the multiplication of unicellular organisms, where uninuclear cells combine within a common (shared) cellular membrane to form a multinuclear organism.

In this paper, the abstract complete graphs $K_4$ of the planar tensegrity cells represent the cells that undergo cellular adhesion and fusion to compose tensegrity structures that exhibit homologous properties such as connectivity, number of elements, one dimensional self-stress space and infinitesimal rigidity. However, other properties depend on the realization of complete graphs $K_4$ in the plane, mainly the geometry and typology of the elements which may differ from a cell to another. The cellular multiplication process proposed allows newly generated tensegrity cells to share nodes and members. If shared elements are not removed, the process corresponds to cellular adhesion: cells are stable and function separately. In this case, the number of self-stress states in the structure increases. If any shared elements are removed, the process corresponds to cellular fusion as the cells composing the system function as one entity. In this case, the number of self-stress states remains the same or decreases. Figure 2 illustrates the analogy between biological and tensegrity cell multiplication mechanisms. Changes in the dimension of self-stress space depend on the number of added nodes, shared members, and removed members. A corollary that allows to keep track of the changes in the dimension of self-stress space is presented in Section 4.

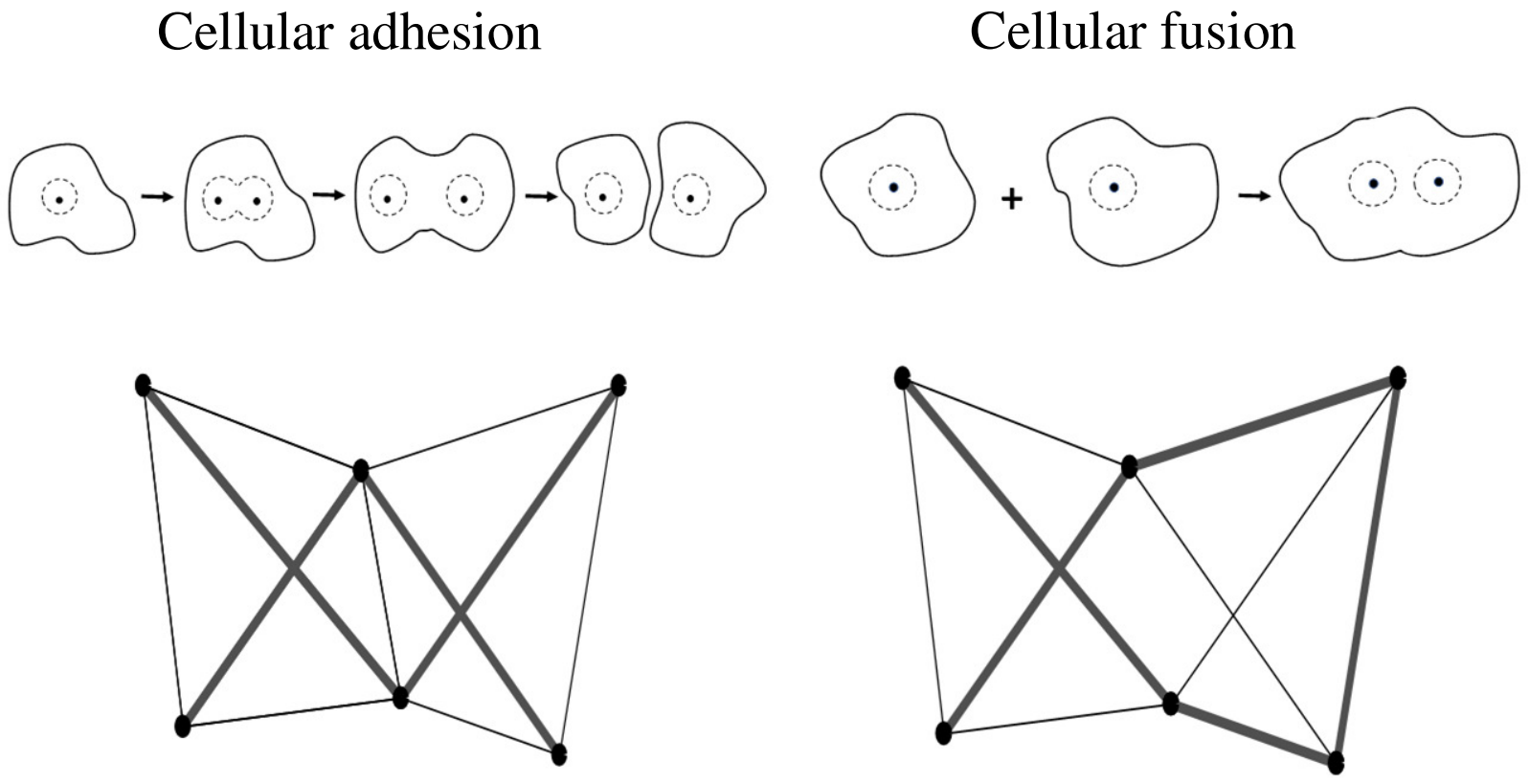

Figure 2: Illustration of the analogy between biological and tensegrity cell multiplication mechanisms.

## 3.2. Cellular multiplication mechanisms

### 3.2.1. Cellular adhesion

In the multiplication process of tensegrity cells, adhesion occurs when a cell is added to the existing structure with all members between the new cell and the existing structure being preserved. Adhesion results thus in an increase of the number of self-stress states in the structure. In this case, the self-stress space is augmented, and the member typology depends only on designer's assignment of the self-stress in the cells. Mathematically, the adhesion of tensegrity structures corresponds to the gluing operation in graph theory. Graph $G=(V,E)$ is the result of the gluing operation of subgraphs $H_1=(V_1,E_1)$ and $H_2=(V_2,E_2)$ if the sets of vertices $V_1, V_2 \subseteq V$ are such that $V_1 \cap V_2 \neq \emptyset$ and $V_1 \cup V_2 = V$ and $E_1$, $E_2$ are the set of edges of $G$ induced by $V_1$, $V_2$. According to Whitely [47], if $H_1$ and $H_2$ are both rigid in $\mathbb{R}^d$ and $| V_1 \cap V_2 | \geq d$ then $G$ will also be rigid in $\mathbb{R}^d$. In this paper, only cases where cells share at least two vertices are treated. Thus, the resulting tensegrity structures from the adhesion mechanism are always rigid, since the composing structures are rigid and they share two nodes or more.

### 3.2.2. Cellular fusion

Fusion occurs if, after the adhesion, one or multiple edges between the added cell and the existing structure are removed. Removed edges can be thought of as members with no self-stress. Consequently, member removal can be achieved by setting the self-stress in the new cell such that the forces in the members to be removed cancel out. Removing one edge is always possible and can be done by adjusting the self-stress in the cell being added. In this case the resulting structure is always rigid. However, for the removal of multiple edges, the new cell has to share at least three nodes with the existing structure. In the case where the number of shared nodes is exactly three, the position of the fourth node defined by coordinates x and y depends on the number of edges being removed. Assuming that all nodes are in general position, the position of the fourth node is determined by the following rules:

- If only one member is being removed, then the added node can take any position. Figure 3 illustrates the fusion between cells *ABCD* and *ABCE* with the configuration of the cells not changing when edge *(B,C)* is removed.

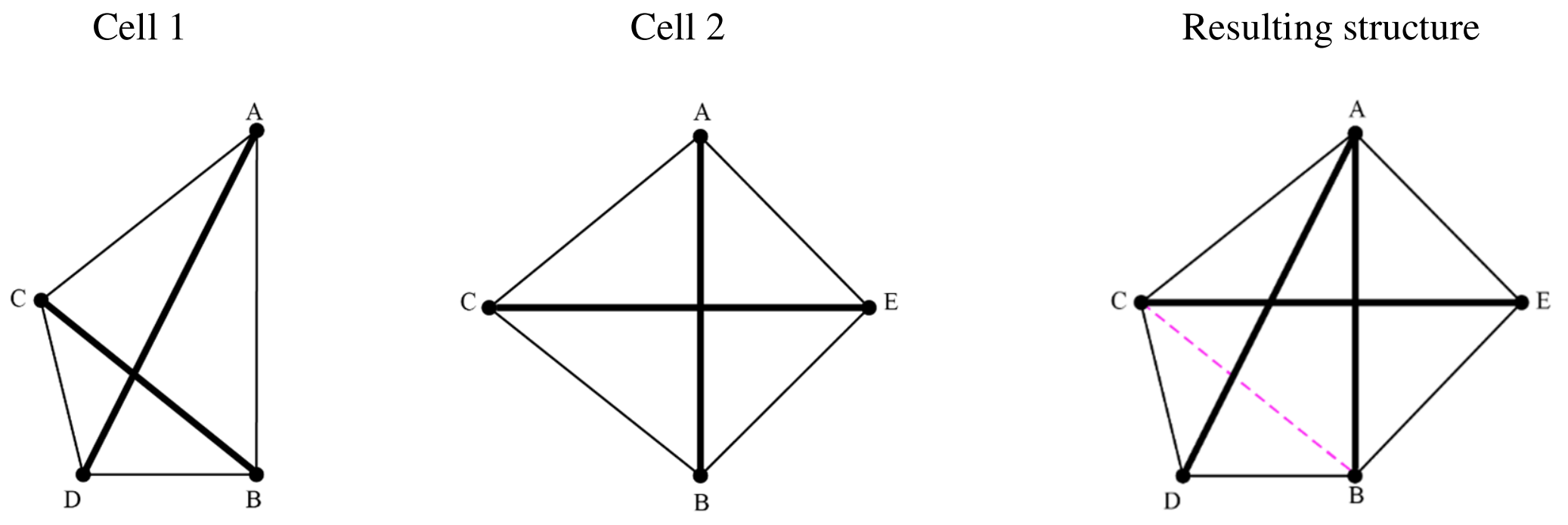

Figure 3: Illustration of cellular fusion between two cells with the removal of an edge.

- If two members are being removed, the self-stress in those two members must be canceled. Without loss of generality, let *A*, *B* and *C* be the shared nodes between cells *ABCD* and *ABCE*. Figure 4 illustrates how the removal of edges *(A,B)* and *(B,C)* described by the dashed lines affects the geometry of the resulting structure (node *E* has to lie on the line defined by nodes *B* and *D* in order for the structure to be in equilibrium).

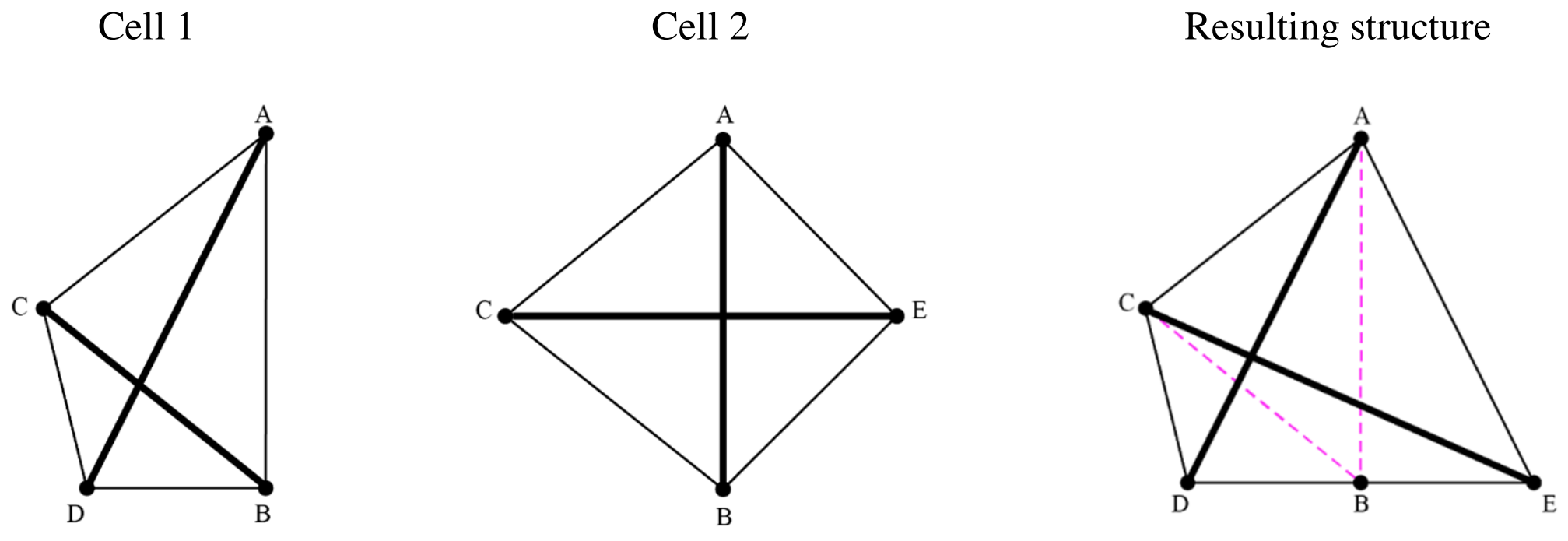

Figure 4: Illustration of cellular fusion between two cells with the removal of two edges.

Assuming the removed members are *(A,B)* and *(B,C)*, and the corresponding self-stress coefficients in those members at the existing structure are $t_1$ and $t_2$, the position of node *D(x,y)* is determined by:

$$\begin{cases} w_1 = \alpha t_1 \\ w_2 = w_1 \dfrac{\mathrm{f}(\mathrm{A},\mathrm{B},(\mathrm{x},\mathrm{y}))}{\mathrm{f}(\mathrm{B},\mathrm{C},(\mathrm{x},\mathrm{y}))} = \alpha t_2 \end{cases} \tag{14}$$

Assuming $\alpha$ is a known coefficient, Equation (14) is a second-degree linear system of three variables ($x$, $y$ and $w_1$). The fourth node lies on the line given by Equation (15) obtained through the removal of $w_1$:

$$\left(-\begin{vmatrix}1 & a_2\\1 & b_2\end{vmatrix}+\frac{t_2}{t_1}\begin{vmatrix}1 & b_2\\1 & c_2\end{vmatrix}\right)x+\left(\begin{vmatrix}1 & a_1\\1 & b_1\end{vmatrix}-\frac{t_2}{t_1}\begin{vmatrix}1 & b_1\\1 & c_1\end{vmatrix}\right)y=\left(-\begin{vmatrix}a_1 & a_2\\b_1 & b_2\end{vmatrix}+\frac{t_2}{t_1}\begin{vmatrix}b_1 & b_2\\c_1 & c_2\end{vmatrix}\right) \tag{15}$$

The position of the fourth node can be determined either by fixing one of the coordinates and using Equation (15), or by fixing $w_1$ and identifying the intersection of Equations (14) and (15).

- If three members are being removed (assuming member *(A,C)* is being removed in addition to members *(A,B)* and *(B,C)*), a third independent linear equation is added to system (14):

$$w_1 \frac{\mathrm{f}(\mathrm{A},\mathrm{B},(\mathrm{x},\mathrm{y}))}{\mathrm{f}(\mathrm{A},\mathrm{C},(\mathrm{x},\mathrm{y}))} = -t_3 \tag{16}$$

Using Equation (16) and one equation from (14) a new line can be defined with its intersection with Equation (15) defining the position of fourth node.

It is important to note, though, that sometimes the removal of edges depends on the configuration (geometry) of the problem, especially when the cell and the existing structure share all four nodes. In this case, removing one edge may result in the removal of multiple edges.

## 4. Changes in the self-stress space during cellular multiplication

### 4.1. Corollary for the number of self-stress states

Cellular multiplication reflects the reverse process of the tensegrity decomposition proposed by de Guzmán and Orden [4]. A corollary to Proposition II given by Fernández and Orden [40], which allows to combinatorically calculate the number of self-stress states by decomposing tensegrity structures into cells, is thus proposed:

**Proposition III.** Let d be the dimension of the workspace. Let $G_i$ and $G_{i+1}$ be the abstract underlying graphs of the tensegrity structures obtained through cellular multiplication at steps *i* and *i+1*. Let $B_i$ and $B_{i+1}$ be their Laman Bounds, respectively. Let $e_i$ be the change in the number of edges between $G_i$ and $G_{i+1}$, and $v_i$ be the change in the number of nodes. $e_i$ and $v_i$ can take positive or negative values depending on whether edges or nodes are being added or removed from the structure. Assuming $G_{i+1}$ is generically rigid, the dimension of the self-stress space *W* in the case of adhesion is given by:

$$\dim(W) = B_{i+1} = B_i + e_i - d * v_i$$

The corollary allows to keep track of the changes in the number of self-stress states during adhesion and fusion, and can be easily proved using the definition of the Laman Bounds $B_i$ and $B_{i+1}$ (Equation 5) and relating the set of edges $E_{i+1}$ to $E_i$ and the set of nodes $V_{i+1}$ to $V_i$. For the planar case, Proposition III resumes in $\Delta(\dim(W)) = e_i - 2v_i$.

### 4.2. Identification of the self-stress space

For planar tensegrity structures, the number of added or removed states can take any integer value from 0 to 6. Figure 5 illustrates examples of changes in the dimension of the self-stress space for common configurations. Although one cell is generated in each step and each cell has only one stable self-stress state, it is shown that depending on the change in the number of nodes $v_i$ and the change in the number of edges $e_i$, the dimension of the self-stress space in the structure between generation steps $i$ and $i+1$ might increase, decrease or remain the same. Since tensegrity cells have one self-stress state, changes in the dimension of the self-stress space larger than the unit are evidence of interactions between the added cell and the existing tensegrity structure.

| Structure at step *i* | Structure at step *i+1* | Self-stress space dimension analysis |
|---|---|---|
| a) 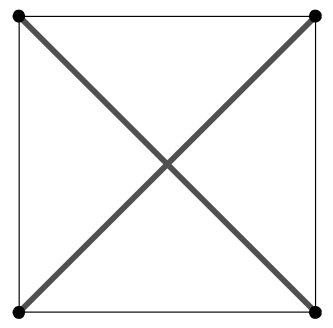 | 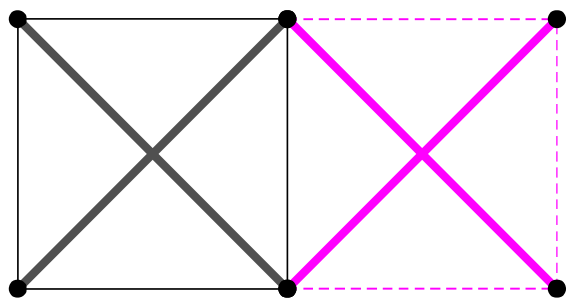 | • $v_i$=2<br>• $e_i$=5<br>⇨ $\Delta(\dim(W)) = e_i - 2v_i = 1$ |
| b) 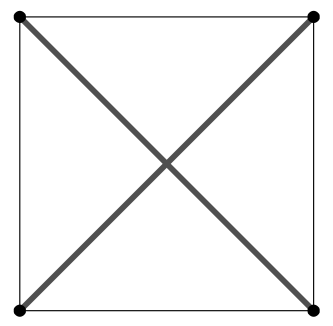 | 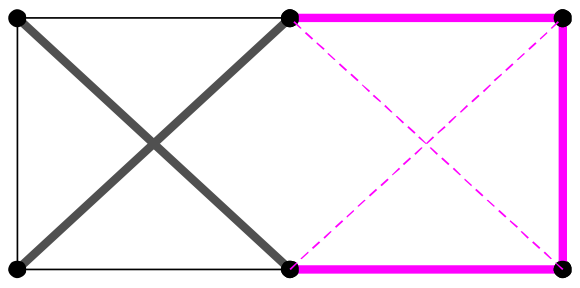 | • $v_i$=2<br>• $e_i$=4<br>⇨ $\Delta(\dim(W)) = e_i - 2v_i = 0$ |
| c) 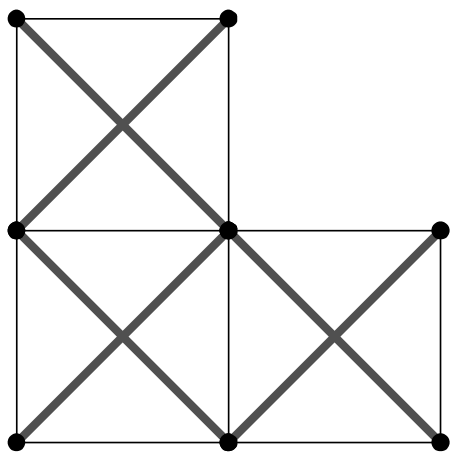 | 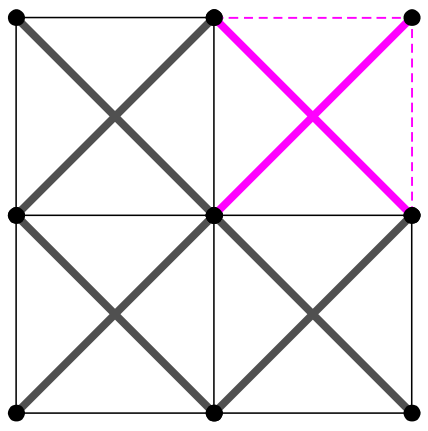 | • $v_i$=1<br>• $e_i$=4<br>⇨ $\Delta(\dim(W)) = e_i - 2v_i = 2$ |
| d) 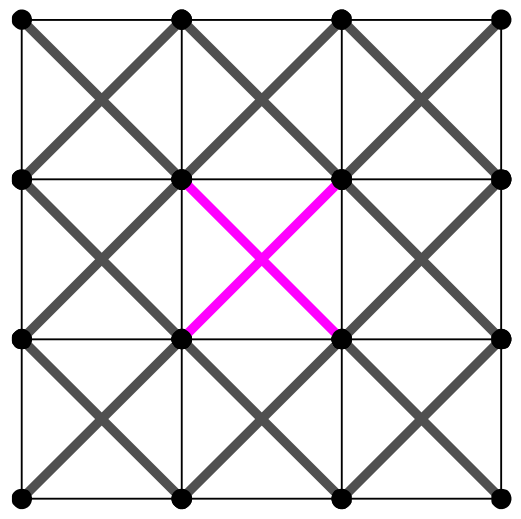 | 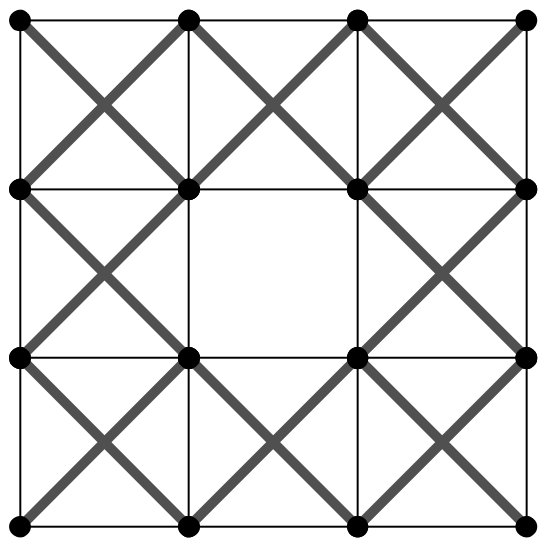 | • $v_i$=0<br>• $e_i$=-2<br>⇨ $\Delta(\dim(W)) = e_i - 2v_i = -2$ |
| e) 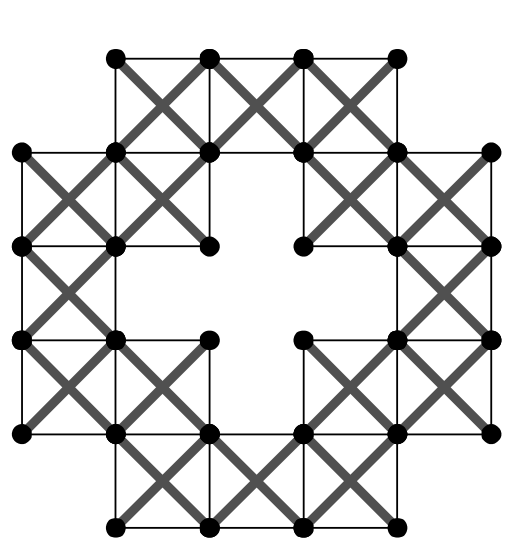 | 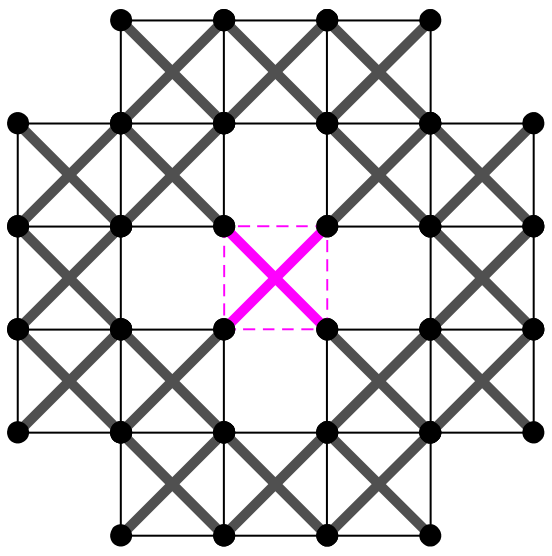 | • $v_i$=0<br>• $e_i$=6<br>⇨ $\Delta(\dim(W)) = e_i - 2v_i = 6$ |

Figure 5: Examples of changes in the dimension of the self-stress space Δ(dim($W$)) during the addition or removal of a cell in various structures ($v_i$: change in the number of nodes, $e_i$: the change in the number of edges).

A basis for the self-stress space of the structure at step $i+1$ can be constructed by calculating its dimension and completing the basis of the structure at step $i$ with any missing vectors. Missing vectors must describe the self-stress state in the added cell, as well as form with the existing basis (basis of the self-stress space of the structure at step $i$) a set of linearly independent vectors of cardinality equal to the dimension of the self-stress space. The simple case of a three-cell structure is treated below to illustrate the process.

The three-cell structure is generated in three steps. In the first step, a Type I tensegrity cell (Figure 6) is generated and its self-stress state is calculated using Equation (5) where $P$ corresponds to the $|V|\times 2$ configuration matrix, Link is a $|E|\times 2$ connectivity matrix containing the nodes of each member and $W$ is the self-stress matrix (here a vector since the states correspond to the columns of matrix $W$ and cells have only one self-stress state). At each step of the process, the dimensions of matrices Link and $W$ are augmented to incorporate members added to the structure and any changes in the number of stable self-stress states.

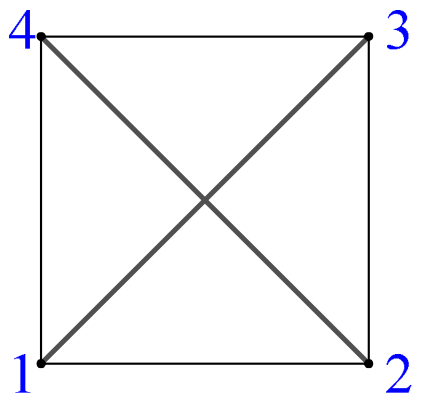

$$P=\begin{bmatrix}0&0\\1&0\\1&1\\0&1\end{bmatrix}\qquad Link=\begin{bmatrix}1&2\\2&3\\3&4\\1&4\\1&3\\2&4\end{bmatrix}\qquad W=\begin{bmatrix}1\\1\\1\\1\\-1\\-1\end{bmatrix}$$

Figure 6: First step in cellular multiplication process of a three-cell tensegrity structure.

In the next step, a second cell is added to the structure (Figure 7). The coordinates of the new cell are specified and employed to calculate the self-stress states. The self-stress matrix W is augmented with zeros to incorporate the new members. The two cells share member {2 3}, and two nodes are added. According to Proposition III the dimension of the self-stress space is equal to 2 with the states corresponding to the self-stress states from the two cells.

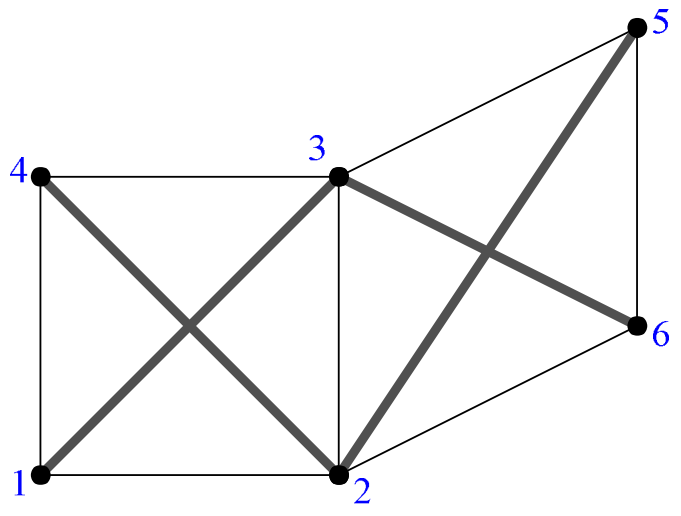

$$P=\begin{bmatrix}0&0\\1&0\\1&1\\0&1\\2&\frac{3}{2}\\2&\frac{1}{2}\end{bmatrix}\qquad Link=\begin{bmatrix}1&2\\2&3\\3&4\\1&4\\1&3\\2&4\\3&5\\5&6\\2&6\\2&5\\3&6\end{bmatrix}\qquad W=\begin{bmatrix}1&0\\1&1\\1&0\\1&0\\-1&0\\-1&0\\0&1\\0&1\\0&1\\0&-1\\0&-1\end{bmatrix}$$

Figure 7: Second step in cellular multiplication process of a three-cell tensegrity structure.

In step 3, the coordinates of the third cell are specified and used to calculate its self-stress state (Figure 8). The third cell shares two members with the existing structure (members {3 4} and {3 5}). Only one new node is added to the structure. The dimension of the self-stress space calculated using Proposition III is equal to four. The addition of the third cell results thus into two additional self-stress states in the structure: one from the newly formed cell and one from its interaction with the existing structure. To find this additional fourth state, it is sufficient to identify a stable sub-structure with a single self-stress state composed of elements from the different cells but without using all the elements of a given cell. In this paper, such a sub-structure is named virtual cell as it is the result of the interaction among cells. The virtual cell for the three-cell structure is illustrated in the inset of Figure 8.

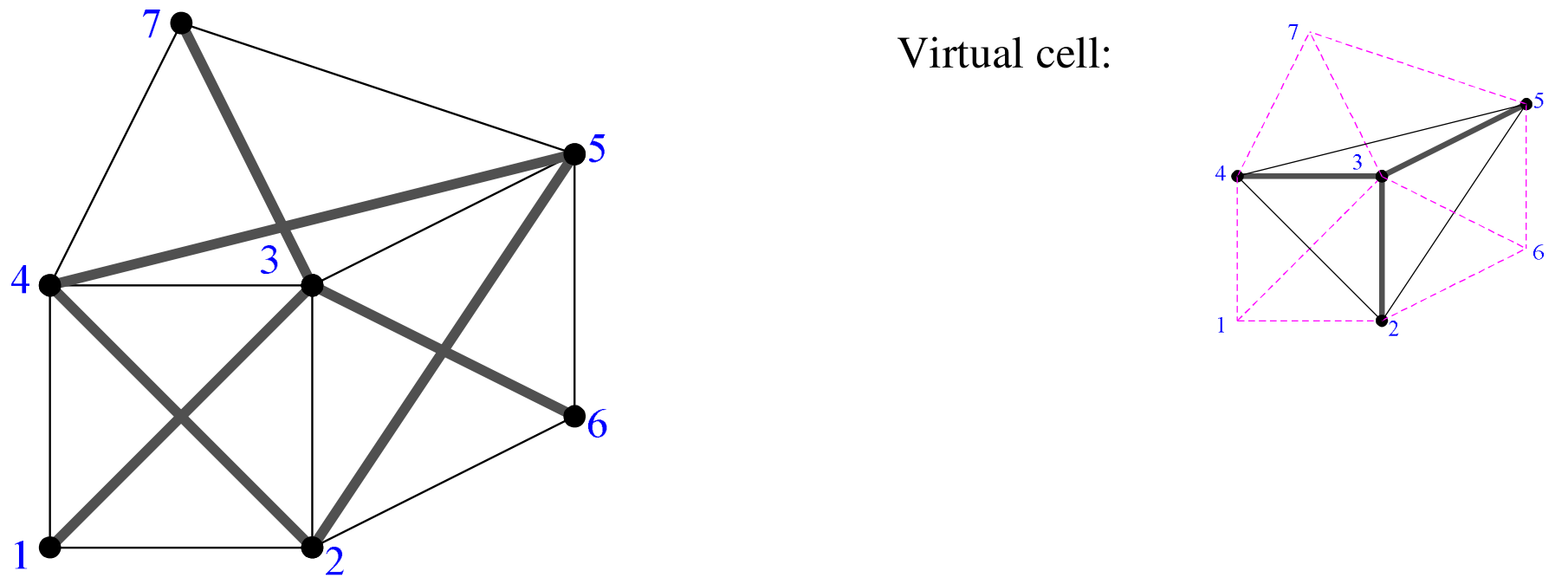

$$P=\begin{bmatrix} 0 & 0 \\ 1 & 0 \\ 1 & 1 \\ 0 & 1 \\ 2 & \frac{3}{2} \\ 2 & \frac{1}{2} \\ \frac{1}{2} & 2 \end{bmatrix} \qquad Link=\begin{bmatrix} 1 & 2 \\ 2 & 3 \\ 3 & 4 \\ 1 & 4 \\ 1 & 3 \\ 2 & 4 \\ 3 & 5 \\ 5 & 6 \\ 2 & 6 \\ 2 & 5 \\ 3 & 6 \\ 5 & 7 \\ 4 & 7 \\ 3 & 7 \\ 4 & 5 \end{bmatrix} \qquad W=\begin{bmatrix} 1 & 0 & 0 & 0 \\ 1 & 1 & 0 & -3.5 \\ 1 & 0 & 1 & -2.3333 \\ 1 & 0 & 0 & 0 \\ -1 & 0 & 0 & 0 \\ -1 & 0 & 0 & 1 \\ 0 & 1 & 0.8 & -2.3333 \\ 0 & 1 & 0 & 0 \\ 0 & 1 & 0 & 0 \\ 0 & -1 & 0 & 1 \\ 0 & -1 & 0 & 0 \\ 0 & 0 & 0.2286 & 0 \\ 0 & 0 & 0.2857 & 0 \\ 0 & 0 & -0.5714 & 0 \\ 0 & 0 & -0.4 & 0.6667 \end{bmatrix}$$

Figure 8: Third step in cellular multiplication process of a three-cell tensegrity structure. The virtual cell, illustrated in the highlight, results in an additional fourth self-stress state in the structure.

To show that a set of vectors defined this way is linearly independent, the null linear combination of the self-stress states is considered. The combination coefficients can be shown to equal zero since the equations corresponding to the elements at the boundary of the structure and which belong to only one cell are of the form $\alpha_j w_{ij}=0$, where $w_{ij}$ is different than zero by the definition of non-trivial self-stress of a cell defined in Section 2.3. This shows that all the coefficients corresponding to the boundary cells are zero. If the boundary cells are removed and the remaining structure is considered, the same argument holds for the boundary cells of the new structure. The process can be iterated until all the coefficients are found to be zero, which proves that the self-stress states are linearly independent.

Virtual cells are collateral stable sub-structures that arise from the interaction between cells and are evoked to complete the self-stress space. It should be noted that the designer is not bound to the proposed method; other methods can be pursued for the construction of the self-stress space resulting in other basis that describe the self-stress space. One can decide to find states that respect the typology of the members which are referred to as conform self-stress states. This can be achieved by multiplying the self-stress matrix $W$ with an invertible matrix $T$ where the coefficients are carefully chosen so that the result force-densities agree with the type of elements as shown below:

$$W_{conform} = W \times T = \begin{bmatrix} 1 & 0 & 0 & 0 \\ 1 & 1 & 0 & -3.5 \\ 1 & 0 & 1 & -2.3333 \\ 1 & 0 & 0 & 0 \\ -1 & 0 & 0 & 0 \\ -1 & 0 & 0 & 1 \\ 0 & 1 & 0.8 & -2.3333 \\ 0 & 1 & 0 & 0 \\ 0 & 1 & 0 & 0 \\ 0 & -1 & 0 & 1 \\ 0 & -1 & 0 & 0 \\ 0 & 0 & 0.2286 & 0 \\ 0 & 0 & 0.2857 & 0 \\ 0 & 0 & -0.5714 & 0 \\ 0 & 0 & -0.4 & 0.6667 \end{bmatrix} \begin{bmatrix} 1 & 1 & 1 & 1 \\ 1 & 0.5 & 1 & 1 \\ 1 & 1 & 0.5 & 1 \\ -1 & -1 & -1 & -0.5 \end{bmatrix} = \begin{bmatrix} 1 & 1 & 1 & 1 \\ 5.5 & 5 & 5.5 & 3.75 \\ 4.3333 & 4.3333 & 3.8333 & 3.1666 \\ 1 & 1 & 1 & 1 \\ -1 & -1 & -1 & -1 \\ -2 & -2 & -2 & -1.5 \\ 4.1333 & 3.6333 & 3.7333 & 2.9667 \\ 1 & 0.5 & 1 & 1 \\ 1 & 0.5 & 1 & 1 \\ -2 & -1.5 & -2 & -1.5 \\ -1 & -0.5 & -1 & -1 \\ 0.2286 & 0.2286 & 0.1143 & 0.2286 \\ 0.2857 & 0.2857 & 0.1429 & 0.2857 \\ -0.5714 & -0.5714 & -0.2857 & -0.5714 \\ -1.0667 & -1.0667 & -0.8667 & -0.7333 \end{bmatrix} \quad (17)$$

The example illustrates the methodology of tensegrity cellular multiplication. Figure 9 presents the flowchart of operations performed during the cellular multiplication process. The process starts with a tensegrity cell and a known self-stress state. Adhesion occurs then, with the addition of new cells to the existing system until the desired tensegrity structure is formed. Every time a cell is added, a search for virtual cells is initiated in order to correctly update the basis describing the self-stress space. Fusion can occur at any time during the process. If fusion occurs, element removal will take place affecting also the self-stress basis. The complexity of the method resides in the identification of the virtual cells and the construction of the self-stress basis. Furthermore, identifying the self-stress states of virtual cells can be challenging in some cases due to the combinatorial nature of the problem. Although it might not be of great importance to a designer who is only interested in finding a planar tensegrity structure in accordance with geometrical constraints (shape, number of elements, symmetry), identifying virtual cells is crucial for describing the self-stress space of a tensegrity structure.

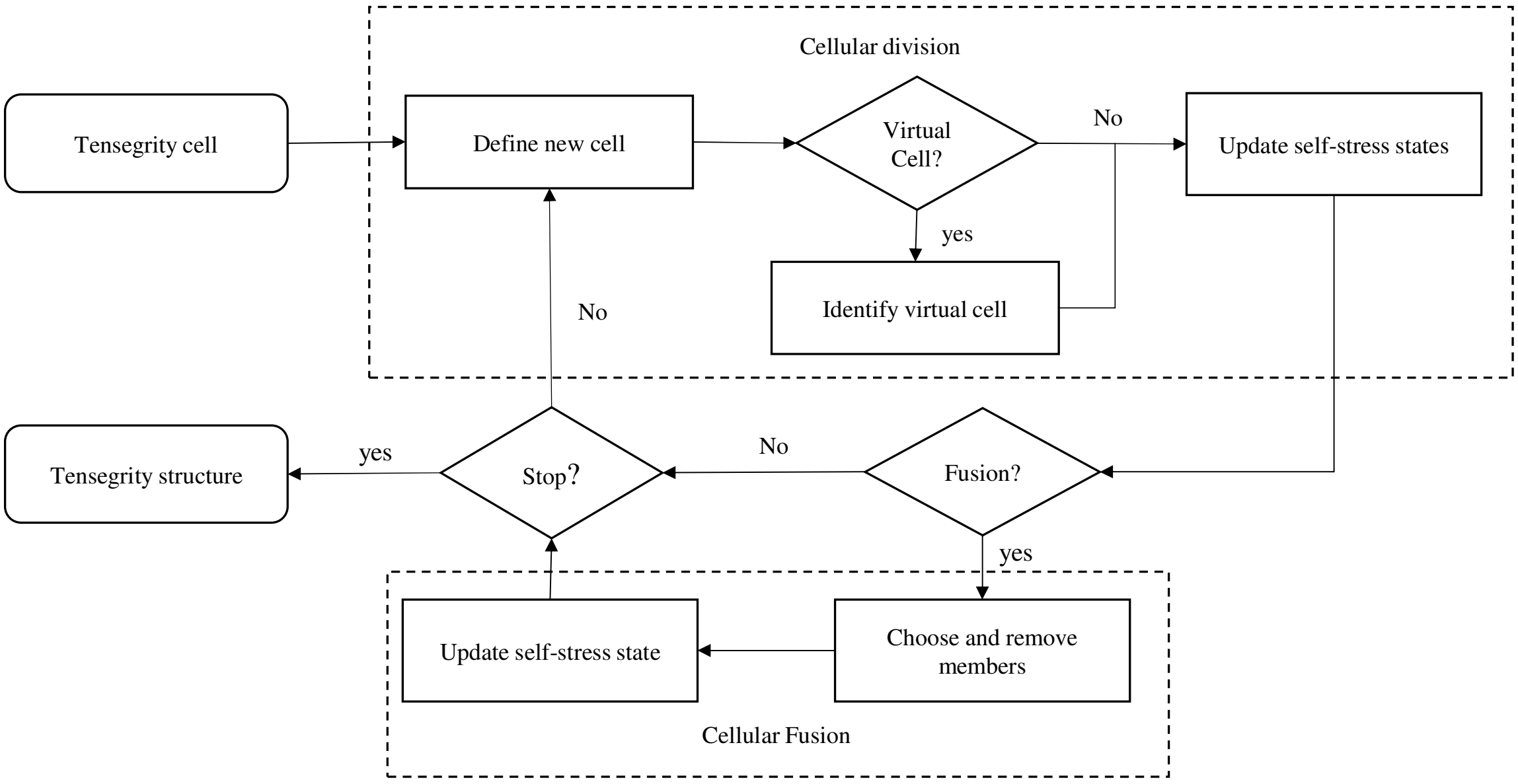

Figure 9: Flow chart of the cellular multiplication process.

## 5. Virtual cells and the interaction among tensegrity cells

### 5.1. Characterization of the virtual cells

In this section, a method is proposed for the identification of virtual cells. Using cells as a basis, a tensegrity structure can be modeled as a weighted multigraph $G_c=(E_c,V_c,W_c)$, where the set of vertices $V_c$ represents the cells, and the set of edges $E_c$ represents the shared edges between the cells. Figure 10 shows the weighted multigraph for the three-cell structure studied in the previous section. $G_c$ is updated at each step by i) adding nodes for both the actual and the virtual cells, and ii) adding the edges labeled by their shared members. Weight values are set to zero if the shared edge is removed, or one if the edge is part of the final structure. Keeping track of the history of the updates in the structure facilitates the search process for the virtual cells, which requires traversing the configuration graph $G$ and the cells graph $G_c$, as virtual cells do not have a fixed configuration (pattern).

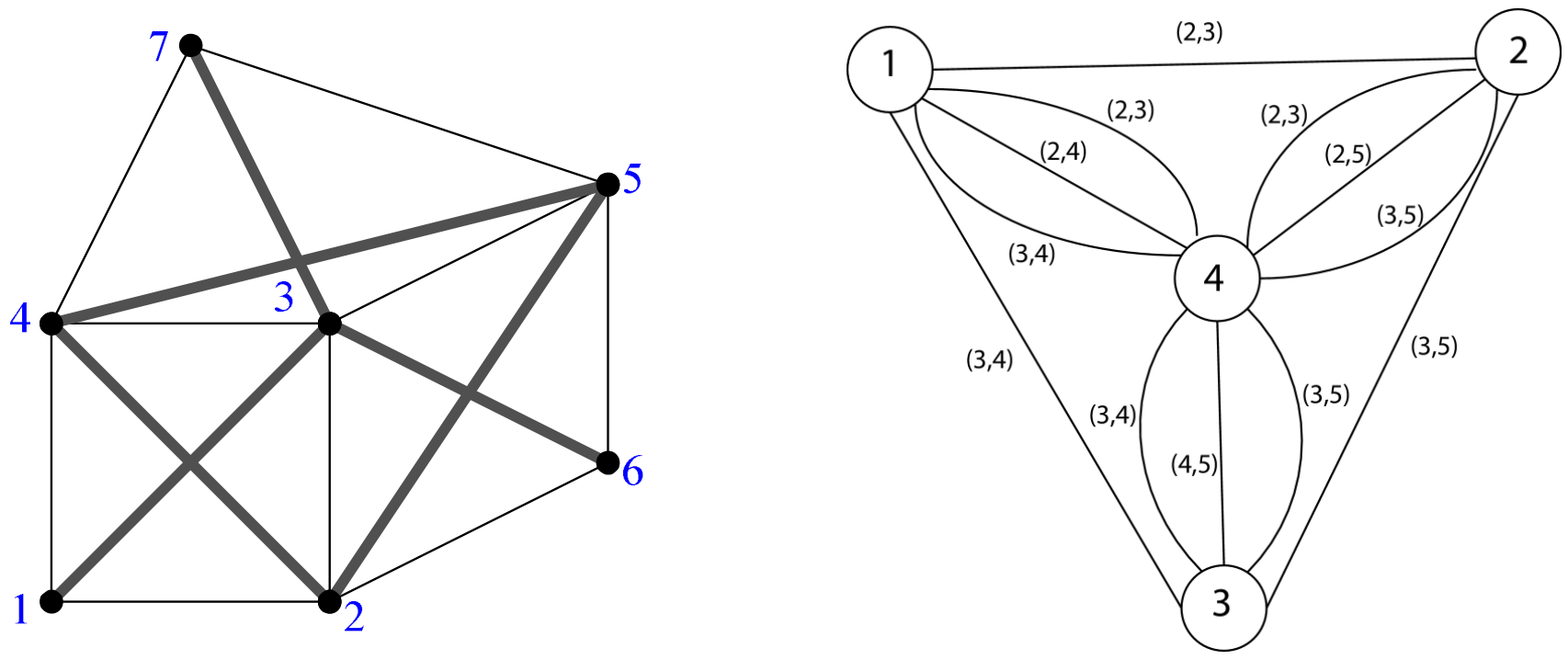

Figure 10: Illustration of a three-cell tensegrity structure and its weighted multigraph.

If the cellular multiplication is performed using the adhesion only (no edge removal occurring throughout the cell multiplication), virtual cells can be considered in the form of a wheel structure. Wheel structures are composed by several peripheral nodes forming a cycle (a tension envelope) and a central node connected to all the peripheral nodes through the compression members (Figure 11). As proved below, wheel structures correspond to tensegrity structures with one self-stress state.

> **Proof.** The base case for the proof is the Type II tensegrity cell which corresponds to a wheel structure of three boundary nodes. Any wheel structure of k nodes on the boundary can be obtained from a wheel structure of k-1 nodes through the addition of a new cell that shares two adjacent nodes and the central node, and the removal of the edge linking the two adjacent nodes. Proposition III shows that the number of self-stress states does not change in this case. Suppose that for all $i \geq 3$ up to $n$, the structures having $i$ nodes at the boundary and obtained through the previous construction step are planar tensegrities and they have one self-stress state. Applying the same construction process to a wheel structure with $n$ boundary nodes will result in another wheel structure with $n+1$ boundary nodes and which has one self-stress state. This concludes the proof, as the property will hold for any wheel structure with $n \geq 3$ boundary nodes.

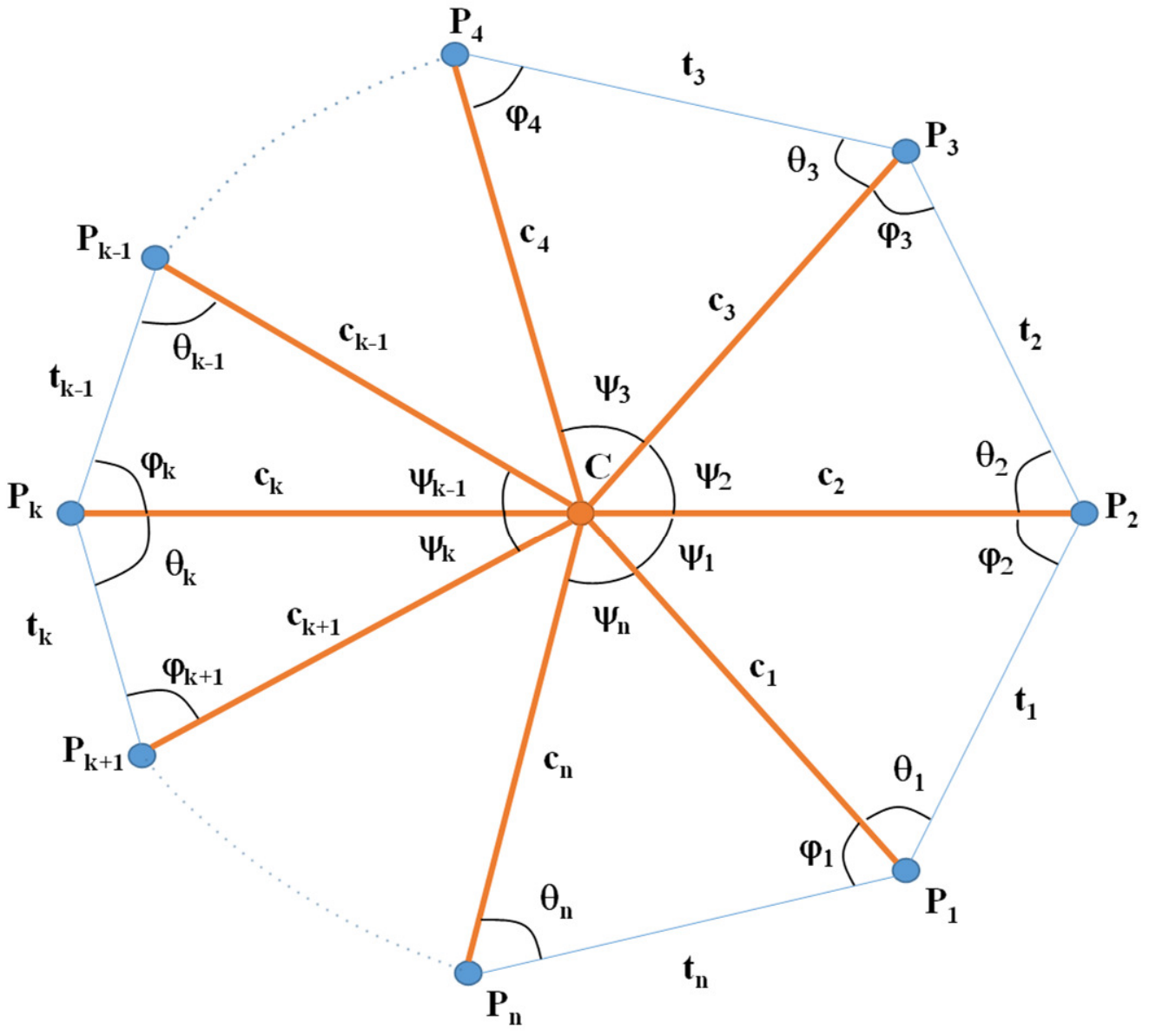

Figure 11: Illustration of a tensegrity-wheel structure.

If fusion is employed during the cellular multiplication process, the resulting structure will be missing one or more edges to complete the required wheel pattern. In this case, the virtual cell will not be in the form of a wheel structure. The substructure must be completed with one or more cells that cancel out the force of the missing members. Consequently, the form of the virtual cell may vary according to the configuration. The wheel pattern can be used for structures that are constructed with cellular adhesion and a limited number of fusion steps. This ensures that a wheel pattern exists or it can be complemented by existing cells to make it stable. However, in general the existence of the wheel pattern is not guaranteed, especially when the structure develops around openings (empty spaces). In this case, a search routine that allows to find a unicellular organism (a minimally rigid tensegrity structure) is implemented to complete the self-stress space with the missing state(s).

### 5.2. Identification of virtual cells

If the tensegrity structure being generated has no openings (empty spaces), virtual cells can be identified using the wheel pattern. Figure 12 presents the flowchart of a search routine based on the wheel pattern. The routine uses as input the configuration graph $G$ and the cells graph $G_c$. It starts by selecting the node with the highest degree of edges in the graph $G$. The immediate neighboring nodes are then selected, and the subgraph induced is isolated. Nodes belonging to only one cell and not shared by multiple cells are removed from the subgraph along with their incident edges. If the subgraph corresponds to a tensegrity wheel structure, then the virtual cell has been detected and its self-stress state is calculated. If this is not the case, removed edges are progressively compensated with removed cells until a stable tensegrity structure is formed and then its self-stress state is calculated.

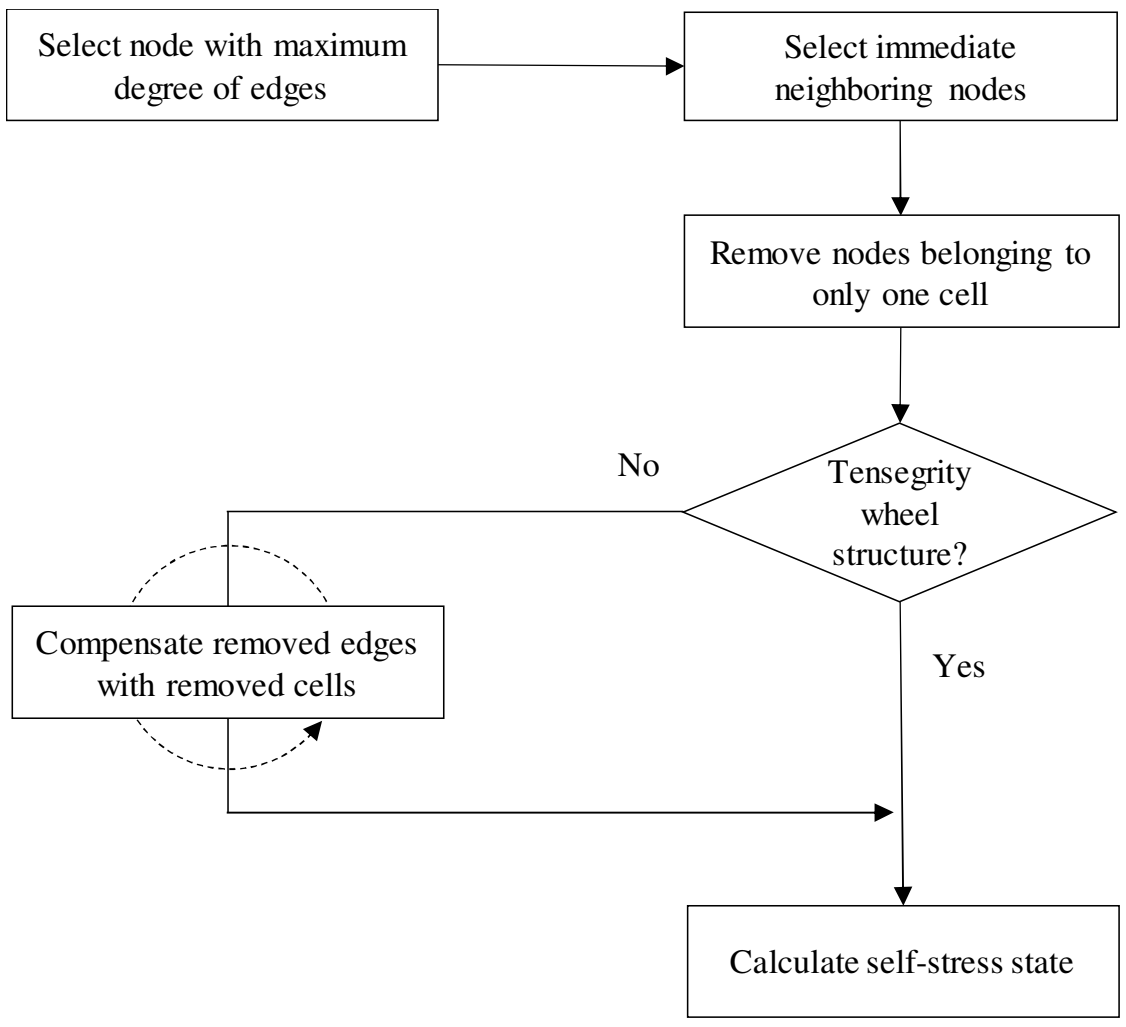

Figure 12: Flowchart of the virtual cell search routine based on the wheel pattern.

The four-cell structures of Figures 13 and 14 are analyzed to exemplify the identification of virtual cells using the proposed search routine. Both structures are generated in four steps, starting from a Type I tensegrity cell with no edge being removed for the structure of Figure 13, while for the structure of Figure 14 the edge (5,6) is removed. In the fourth step of the cellular multiplication process, Proposition III reveals that the number of self-stress states in the structure of Figure 13 is equal to five, with four states coming from the cells composing the structure and one state coming from their interaction. To identify the virtual cell and its self-stress state, node 5 is selected. The immediate neighbors of node 5 are selected and then nodes {1,3,7,9} are removed since they belong to only one cell. The subgraph induced by the remaining nodes is a planar tensegrity wheel.

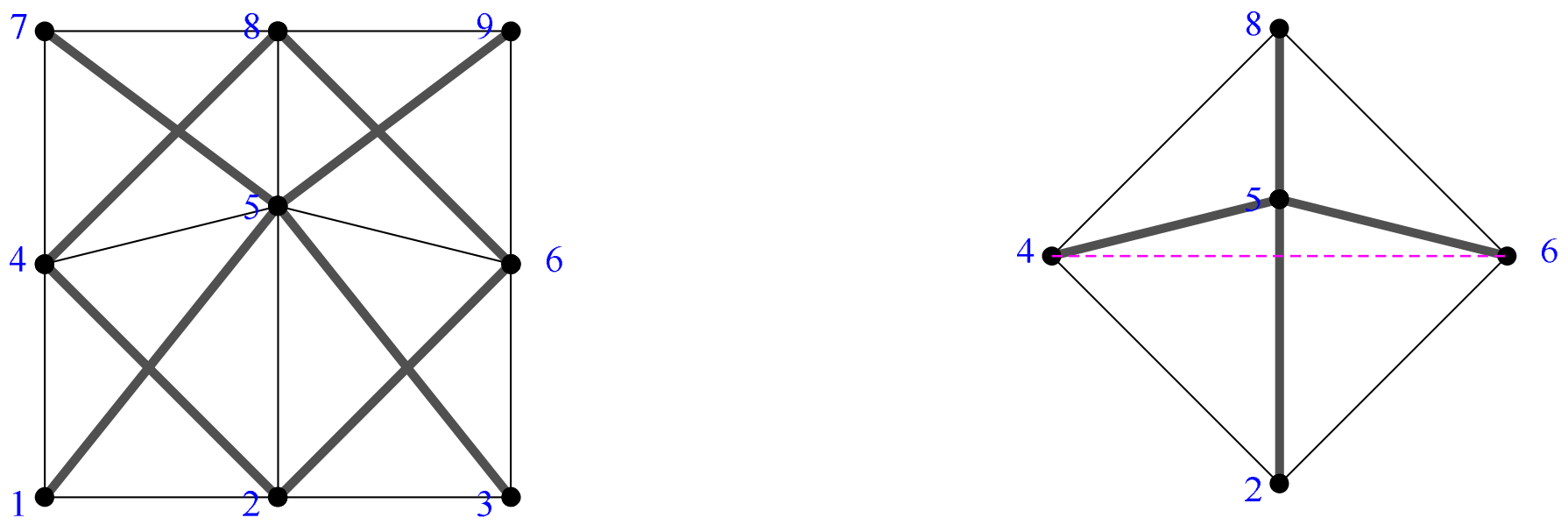

Figure 13: Illustration of a four-cell structure generated with no edge removal and the virtual cell identified.

The structure of Figure 14 is similar to the structure of Figure 13 with the exception of edge (5,6) being removed, which results in a change of typology in the members of the fourth cell. The application of the virtual cell search routine leads again to the selection of node 5 and the substructure induced by nodes {2,4,5,6,8}. However, the resulting substructure is not stable unless the removal of edge (5,6) is balanced by a cell that shares this edge with the substructure. In this case, this cell corresponds to the fourth cell {5,6,8,9}. Adding the fourth cell to the substructure results in a stable tensegrity with one self-stress state, which can be found through the composition of the self-stress state of the wheel structure with the self-stress state of the fourth cell.

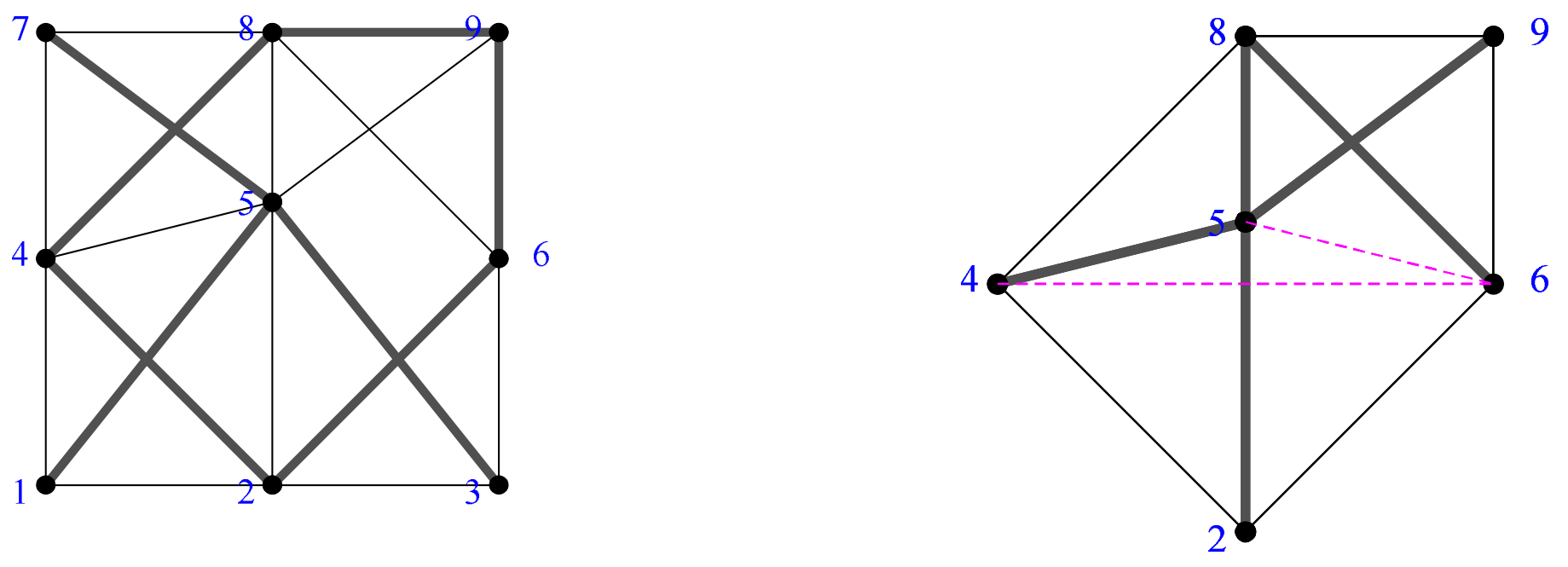

Figure 14: Illustration of a four-cell structure generated with the removal of edge (5,6) and the virtual cell identified.

5.3. Analytical solution of the tensegrity wheel

The wheel structure provides a standardized way in characterizing virtual cell topologies. Therefore, developing an analytical solution for the self-stress of the wheel structures is important for completion of the self-stress basis of a structure. Consider a wheel structure defined by $n$ peripheral nodes denoted $P_k(p_{k1}, p_{k2})$ with $1 \le k \le n$, and a central node $C(c_1, c_2)$. Let $\theta_k, \varphi_k$ and $\psi_k$be the angles as defined in Figure 11. At $P_k$, it can be observed that $\theta_k + \varphi_k \in [0, \pi]$ implying that $\text{sign}(t_k) = -\text{sign}(c_k)$ where $t_k$ is the tension self-stress coefficient (sign($t_k$)=+) in member $P_k P_{k+1}$ and $c_k$ is the compression self-stress coefficient (sign($c_k$)=–) in member $CP_k$. The equilibrium at node $P_k$ gives:

$$t_k \overrightarrow{P_k P_{k+1}} + t_{k-1} \overrightarrow{P_k P_{k-1}} + c_k \overrightarrow{P_k C} = \vec{0} \tag{18}$$

while at node C:

$$\sum_{k=1}^{n} c_k \overrightarrow{CP_k} = \vec{0} \tag{19}$$

A recursive relation between self-stress coefficients $t_k$, $c_k$ and $t_{k-1}$ can be found exploring the equilibrium. Applying the cross-product of the vectors $\overrightarrow{P_k C}$ and $\overrightarrow{P_k P_{k+1}}$ to Equation (18) gives:

$$\left(t_k\overrightarrow{P_kP_{k+1}}\times\overrightarrow{P_kC}+t_{k-1}\overrightarrow{P_kP_{k-1}}\times\overrightarrow{P_kC}\right)\times\overrightarrow{P_kC}=\vec{0}$$

$$\begin{aligned}
&\Rightarrow t_k\overrightarrow{P_kP_{k+1}}\times\overrightarrow{P_kC}+t_{k-1}\overrightarrow{P_kP_{k-1}}\times\overrightarrow{P_kC}+c_k\overrightarrow{P_kC}\times\overrightarrow{P_kC}=\vec{0}\\
&\Rightarrow t_k\overrightarrow{P_kP_{k+1}}\times\overrightarrow{P_kC}=-t_{k-1}\overrightarrow{P_kP_{k-1}}\times\overrightarrow{P_kC}\\
&\Rightarrow |t_k|\left\|\overrightarrow{P_kP_{k+1}}\times\overrightarrow{P_kC}\right\|=|t_{k-1}|\left\|\overrightarrow{P_kP_{k-1}}\times\overrightarrow{P_kC}\right\|\\
&\Rightarrow |t_k|\left\|\overrightarrow{P_kP_{k+1}}\right\|\left\|\overrightarrow{P_kC}\right\|\sin(\theta_k)=|t_{k-1}|\left\|\overrightarrow{P_kP_{k-1}}\right\|\left\|\overrightarrow{P_kC}\right\|\sin(\varphi_k)\\
&\Rightarrow t_k=\frac{\sin(\varphi_k)}{\sin(\theta_k)}\frac{\left\|\overrightarrow{P_kP_{k-1}}\right\|}{\left\|\overrightarrow{P_kP_{k+1}}\right\|}t_{k-1}
\end{aligned} \tag{19}$$

$$\left(t_k\overrightarrow{P_kP_{k+1}}\times\overrightarrow{P_kC}+t_{k-1}\overrightarrow{P_kP_{k-1}}\times\overrightarrow{P_kC}\right)\times\overrightarrow{P_kP_{k-1}}=\vec{0}$$

$$\begin{aligned}
&\Rightarrow t_k\overrightarrow{P_kP_{k+1}}\times\overrightarrow{P_kP_{k-1}}+t_{k-1}\overrightarrow{P_kP_{k-1}}\times\overrightarrow{P_kP_{k-1}}+c_k\overrightarrow{P_kC}\times\overrightarrow{P_kP_{k-1}}=\vec{0}\\
&\Rightarrow c_k\overrightarrow{P_kC}\times\overrightarrow{P_kP_{k-1}}=-t_k\overrightarrow{P_kP_{k+1}}\times\overrightarrow{P_kP_{k-1}}\\
&\Rightarrow |c_k|\left\|\overrightarrow{P_kC}\times\overrightarrow{P_kP_{k-1}}\right\|=|t_k|\left\|\overrightarrow{P_kP_{k+1}}\times\overrightarrow{P_kP_{k-1}}\right\|\\
&\Rightarrow |c_k|\left\|\overrightarrow{P_kC}\right\|\left\|\overrightarrow{P_kP_{k-1}}\right\|\sin(\varphi_k)=|t_{k-1}|\left\|\overrightarrow{P_kP_{k+1}}\right\|\left\|\overrightarrow{P_kP_{k-1}}\right\|\sin(\theta_k+\varphi_k)\\
&\Rightarrow c_k=-\frac{\sin(\theta_k+\varphi_k)}{\sin(\varphi_k)}\frac{\left\|\overrightarrow{P_kP_{k+1}}\right\|}{\left\|\overrightarrow{P_kC}\right\|}t_k
\end{aligned} \tag{20}$$

The components in the self-stress vector $w_{wheel}$ can be rearranged such that the first $n$ components correspond to the tension coefficients $t_k$ and the next $n$ components are the compression coefficients $c_k$. Expressing all coefficients in the self-stress vector $w_{wheel}$ as a function of $t_1$ gives:

$$w_{wheel} = \begin{bmatrix} \left.\begin{matrix} t_1 \\ \frac{\sin(\varphi_2)}{\sin(\theta_2)} \frac{\left\| \overrightarrow{P_2P_1} \right\|}{\left\| \overrightarrow{P_2P_3} \right\|} t_1 \\ \vdots \\ \prod_{i=2}^{k} \frac{\sin(\varphi_i)}{\sin(\theta_i)} \frac{\left\| \overrightarrow{P_iP_{i-1}} \right\|}{\left\| \overrightarrow{P_iP_{i+1}} \right\|} t_1 \\ \vdots \\ \prod_{i=2}^{n} \frac{\sin(\varphi_i)}{\sin(\theta_i)} \frac{\left\| \overrightarrow{P_iP_{i-1}} \right\|}{\left\| \overrightarrow{P_iP_{i+1}} \right\|} t_1 \end{matrix}\right\} t_k \\ \left.\begin{matrix} -\frac{\sin(\theta_1+\varphi_1)}{\sin(\varphi_1)} \frac{\left\| \overrightarrow{P_1P_2} \right\|}{\left\| \overrightarrow{P_1C} \right\|} t_1 \\ -\frac{\sin(\theta_2+\varphi_2)}{\sin(\varphi_2)} \frac{\left\| \overrightarrow{P_2P_3} \right\|}{\left\| \overrightarrow{P_2C} \right\|} t_2 \\ \vdots \\ -\frac{\sin(\theta_k+\varphi_k)}{\sin(\varphi_k)} \frac{\left\| \overrightarrow{P_kP_{k+1}} \right\|}{\left\| \overrightarrow{P_kC} \right\|} t_k \\ \vdots \\ -\frac{\sin(\theta_n+\varphi_n)}{\sin(\varphi_n)} \frac{\left\| \overrightarrow{P_nP_1} \right\|}{\left\| \overrightarrow{P_nC} \right\|} t_n \end{matrix}\right\} c_k \end{bmatrix} \tag{21}$$

For implementation purposes, the components in the self-stress vector of a tensegrity wheel structure can be expressed using the function f (Equation 11) as follows:

$$t_i = \frac{\overrightarrow{OC}\times\overrightarrow{OP_{i-1}} - \overrightarrow{OC}\times\overrightarrow{OP_i} + \overrightarrow{OP_{i-1}}\times\overrightarrow{OP_i}}{\overrightarrow{OC}\times\overrightarrow{OP_i} - \overrightarrow{OC}\times\overrightarrow{OP_{i+1}} + \overrightarrow{OP_i}\times\overrightarrow{OP_{i+1}}} t_{i-1} = \frac{\begin{vmatrix} 1 & c_1 & c_2 \\ 1 & p_{(i-1)1} & p_{(i-1)2} \\ 1 & p_{i1} & p_{i2} \end{vmatrix}}{\begin{vmatrix} 1 & c_1 & c_2 \\ 1 & p_{i1} & p_{i2} \\ 1 & p_{(i+1)1} & p_{(i+1)2} \end{vmatrix}} t_{i-1} = \frac{\mathrm{f}(\mathrm{C}, \mathrm{P}_{i-1}, \mathrm{P}_i)}{\mathrm{f}(\mathrm{C}, \mathrm{P}_i, \mathrm{P}_{i+1})} t_{i-1} \tag{22}$$

$$c_i = -\frac{\overrightarrow{OP_{i-1}}\times\overrightarrow{OP_i} - \overrightarrow{OP_{i-1}}\times\overrightarrow{OP_{i+1}} + \overrightarrow{OP_i}\times\overrightarrow{OP_{i+1}}}{\overrightarrow{OC}\times\overrightarrow{OP_i} - \overrightarrow{OC}\times\overrightarrow{OP_{i+1}} + \overrightarrow{OP_i}\times\overrightarrow{OP_{i+1}}} t_{i-1} = -\frac{\begin{vmatrix} 1 & p_{(i-1)1} & p_{(i-1)2} \\ 1 & p_{i1} & p_{i2} \\ 1 & p_{(i+1)1} & p_{(i+1)2} \end{vmatrix}}{\begin{vmatrix} 1 & c_1 & c_2 \\ 1 & p_{i1} & p_{i2} \\ 1 & p_{(i+1)1} & p_{(i+1)2} \end{vmatrix}} t_{i-1} = -\frac{\begin{vmatrix} 1 & p_{(i-1)1} & p_{(i-1)2} \\ 1 & p_{i1} & p_{i2} \\ 1 & p_{(i+1)1} & p_{(i+1)2} \end{vmatrix}}{\begin{vmatrix} 1 & c_1 & c_2 \\ 1 & p_{(i-1)1} & p_{(i-1)2} \\ 1 & p_{i1} & p_{i2} \end{vmatrix}} t_i = -\frac{\mathrm{f}(\mathrm{P}_{i-1}, \mathrm{P}_i, \mathrm{P}_{i+1})}{\mathrm{f}(\mathrm{C}, \mathrm{P}_{i-1}, \mathrm{P}_i)} t_i \tag{23}$$

Expressing all coefficients in the self-stress vector $w_{wheel}$ as function of $t_1$ leads to:

$$w_{wheel} = \begin{bmatrix} \left.\begin{matrix} t_1 \\ \dfrac{f(C,P_1,P_2)}{f(C,P_2,P_3)} t_1 \\ \vdots \\ \prod_{i=1}^{k-1} \dfrac{f(C,P_{i-1},P_i)}{f(C,P_i,P_{i+1})} t_1 \\ \vdots \\ \dfrac{f(C,P_{n-1},P_n)}{f(C,P_n,P_1)} \prod_{i=1}^{n-1} \dfrac{f(C,P_{i-1},P_i)}{f(C,P_i,P_{i+1})} t_1 \end{matrix}\right\} t_i \\ \left.\begin{matrix} -\dfrac{f(P_n,P_1,P_2)}{f(C,P_n,P_1)} t_1 \\ -\dfrac{f(P_1,P_2,P_3)}{f(C,P_1,P_2)} t_2 \\ \vdots \\ -\dfrac{f(P_{k-1},P_k,P_{k+1})}{f(C,P_{k-1},P_k)} t_k \\ \vdots \\ -\dfrac{f(P_{n-1},P_n,P_{n+1})}{f(C,P_{n-1},P_n)} t_n \end{matrix}\right\} c_i \end{bmatrix} \tag{24}$$

5.4. A general routine for the identification of virtual cells

Employing the wheel pattern provides a direct way of identifying virtual cells in a tensegrity structure as well as an analytical solution to the self-stress space. However, wheel patterns are not found in all cellular multiplication schemes. Therefore, a general routine for the identification of virtual cells is proposed hereby. The routine identifies virtual cells as subgraphs that underlie unicellular organisms: sub-system of a single self-stress state. Once a unicellular organism is identified, its self-stress state is obtained numerically by solving the equilibrium problem (nullspace of the equilibrium matrix $A$). The routine for the identification of virtual cells in a planar tensegrity structure is described below.

Let $s$ be the number of self-stress states at a step $i$ and $p$ the number of cells employed in the generation of a structure until step $i$.

i. Start by removing an edge belonging to one cell only from each cell of the structure. If the edge removed is attached to a node of degree 3 (with three elements connected to it), the removal of the edge may result in the removal of the other two edges attached to that node and thus the removal of the node itself. The removal of an edge, or 3 edges and a node, always results in decreasing the dimension of the self-stress space by a one. The number of self-stress states remaining in the structure is $s$-$p$.

ii. From the remaining structure, select $s$-$p$-$1$ edges to remove, with each edge having both end nodes with a degree equal or larger to 4. The remaining structure will be a unicellular organism with exactly one self-stress state.

iii. Calculate the self-stress state in the unicellular organism by finding the nullspace of its equilibrium matrix. Update the self-stress space of the total structure.

iv. Go back to step (ii) and repeat the process by considering a different set of edges until all virtual cells have been identified.

Figure 15 shows the virtual cells obtained using the proposed routine on a circular tensegrity structure with central opening with the virtual cell candidates identified not corresponding to wheel patterns.

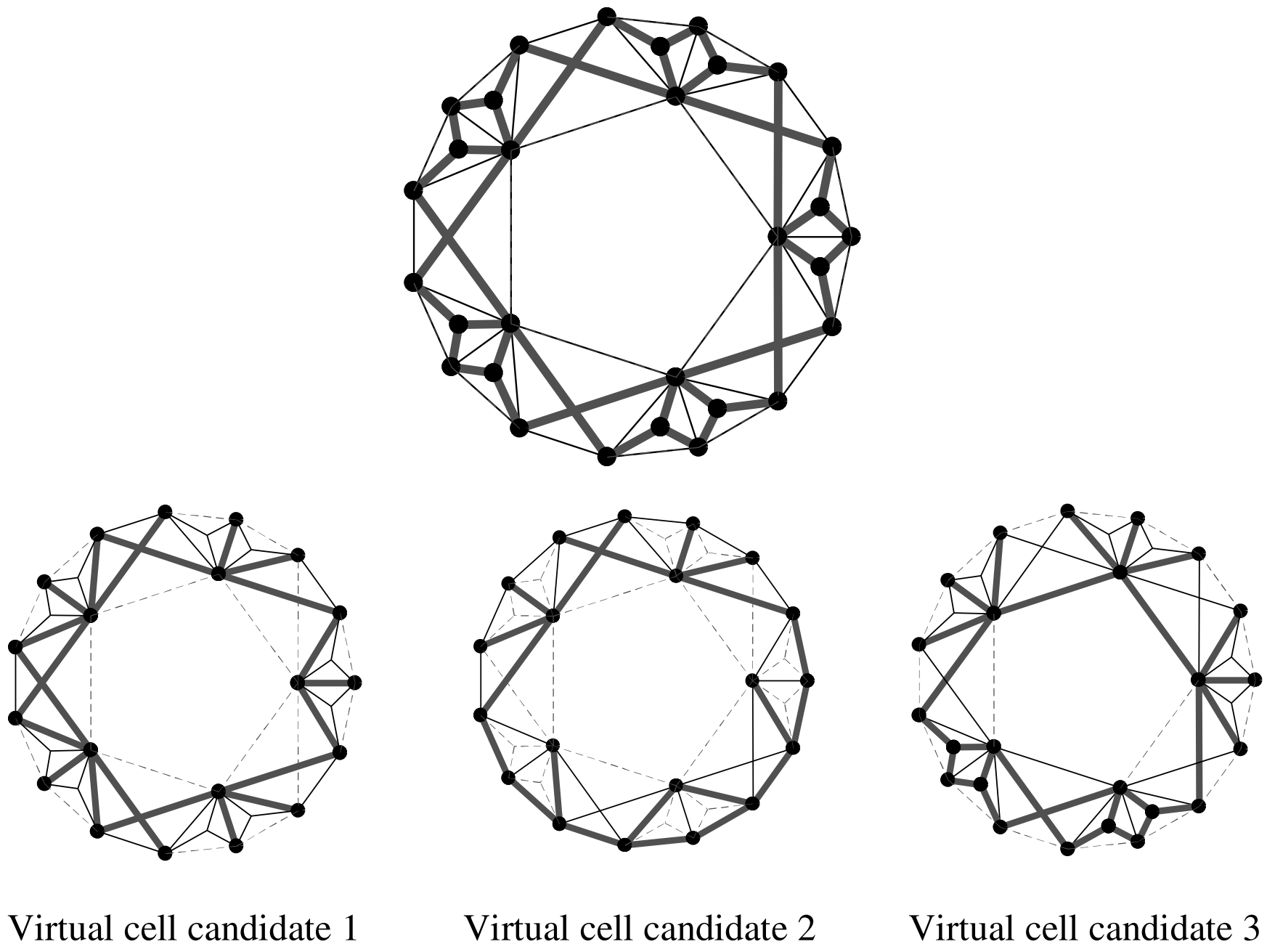

Figure 15: Virtual cells identified for a circular tensegrity structure with central opening (no wheel pattern).

## 6. Examples of tensegrity structures generated with the cellular multiplication method

The cellular multiplication process offers a generation method for planar tensegrity structures where parameters such as the equilibrium shape (form), the number of elements, and the number of self-stress states can be controlled. The proposed method was implemented in Python and employs Networkx, a Python package for graph theory. The method is based on a series of algebraic calculations and characterized by a polynomial time, making it efficient for the generation of large planar tensegrity structures. Two examples of tensegrity structures generated with the cellular multiplication method with predefined shapes and various numbers of self-stress states are presented in this Section to showcase the potential of the method.

In the first example, a circular profile is used to generate planar tensegrity structures with rotational symmetry. Structures are composed of twenty cells (5 Type I cells and 15 Type II cells), with different elements being removed resulting in different numbers of self-stress states (Figure 16). Cellular multiplication principles allow the designer to calculate the number of self-stress states using Proposition III. In simple cases like the structures in Figure 16 a) the number of self-stress states can also be found by visually examining the structure. Virtual cells represent a wheel pattern that can only exist if a node is all surrounded by cells. Thus, the number of self-stress states is given by the number of cells plus the number of nodes that do not lie on the boundary and that are shared by multiple cells minus the number of members being removed.

a)

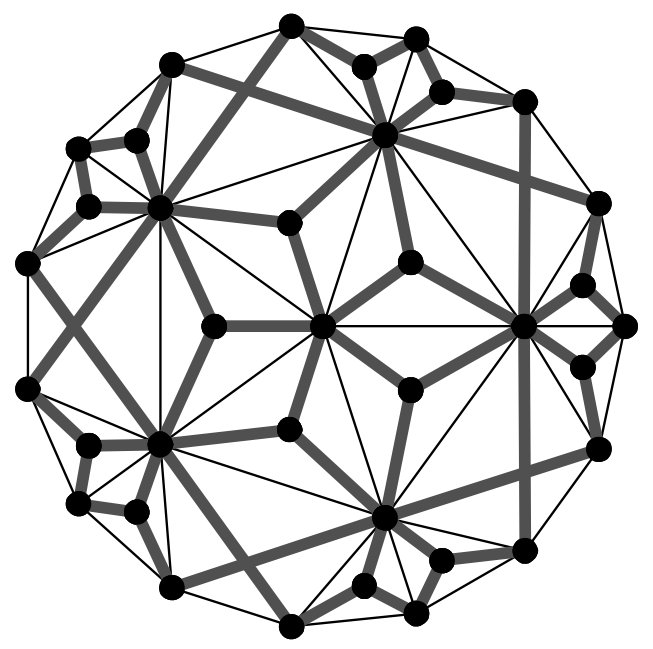

- 20 adhesions
- 0 fusion
- 6 virtual cells

⇨ 26 self-stress states

b)

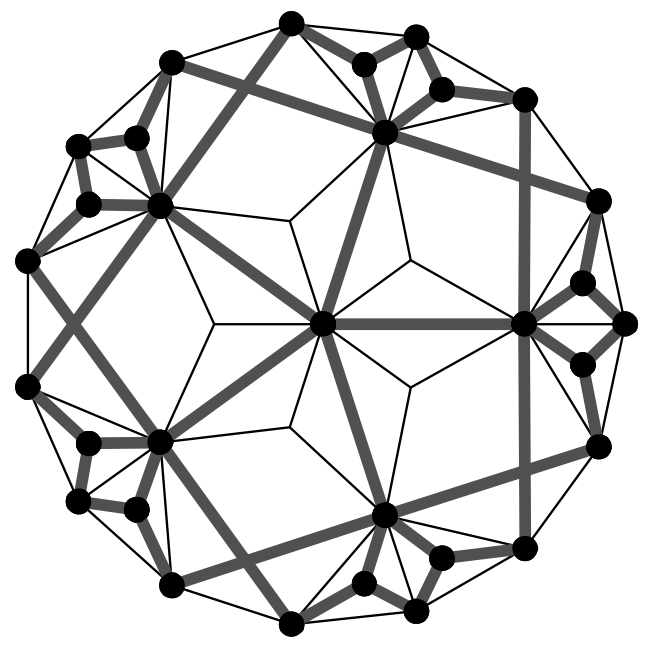

- 20 adhesions
- 5 fusions
- 6 virtual cells

⇨ 21 self-stress states

c)

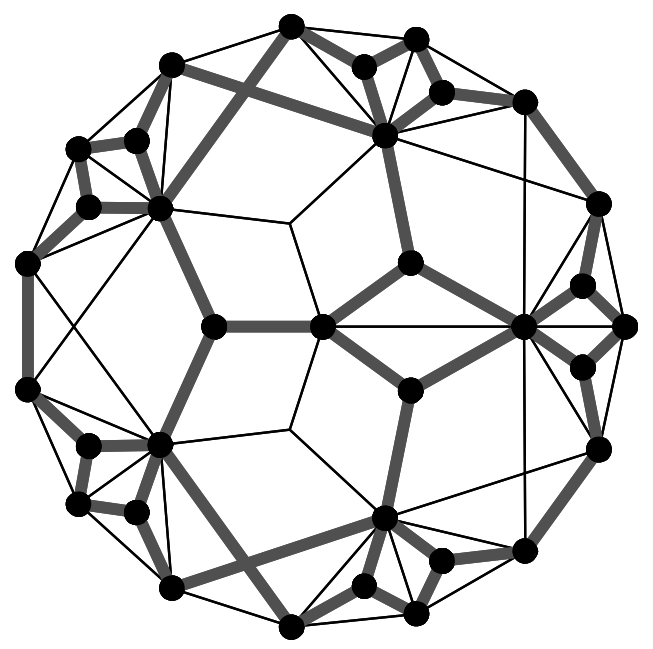

- 20 adhesions
- 9 fusions
- 6 virtual cells

⇨ 17 self-stress states

d)

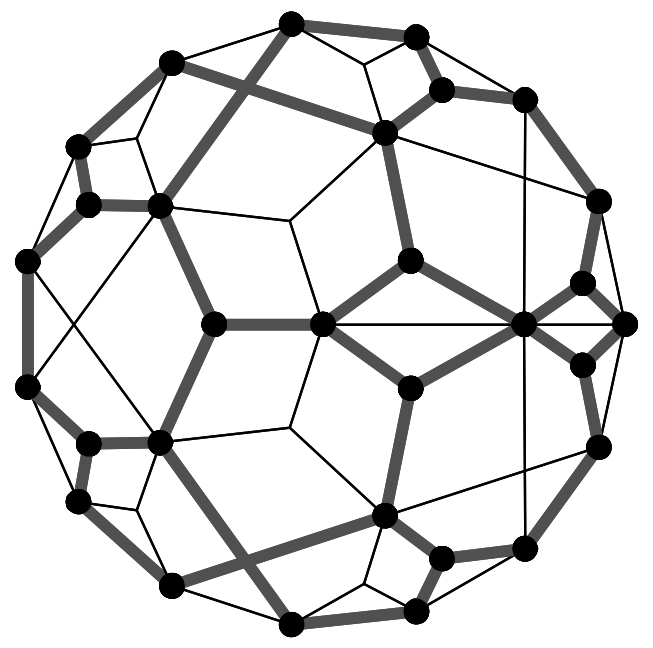

- 20 adhesions
- 23 fusions
- 6 virtual cells

⇨ 3 self-stress states

Figure 16: Examples of circular tensegrity structures generated using the cellular multiplication method and different numbers of self-stress states.

In the second example, the cellular multiplication is employed to generate a large tensegrity structure of elliptical shape and multiple self-stress states. For practical reasons, this is done through an automated implementation of the multiplication mechanisms. The process starts by meshing the desired shape with a quadrilateral or triangular mesh, or a combination of both. There is no constraint on the number of nodes or the shape of the mesh. Nevertheless, the designer can introduce any constraint on the shape, the number, and the size of the mesh. Once the shape has been meshed, the cellular multiplication process can take place. Type I cells can be used in the case of quadrilateral mesh, while Type II cells can be employed in the case of triangular mesh. Other combinations are also possible. Tensegrity cells are stored and accessed in every adhesion step, to define the position of the next cellular adhesion. The process continues until the entire mesh has been populated with tensegrity cells. It should be noted that throughout this process, the designer can modify the multiplication strategy and redirect the process. The main challenge with the automated implementation of the cellular multiplication is the identification of the additional self-stresses resulting from the virtual cells. However, when no members are removed the number of additional self-stress states is equal to the number of nodes shared by multiple cells in the structure.

Figure 17 shows snapshots during the generation process of the large elliptical tensegrity structure. The structure is composed of 70 Type II cells. The elliptical profile and the number of mesh nodes are the only input needed to generate the structure. Nodes in this example are based on a symmetric pattern that covers the area of the ellipse. Delaunay triangulation is then used to triangulate the area, based on the positions of the nodes. Additional nodes are added in the center of the triangles to complete the topology of Type II cells. The assigned profile is then populated using a modified version of the boundary-fill algorithm. The starting cell is chosen randomly and the built structure boundary is set to the neighboring cells of the start cell. In each multiplication step, the boundary is updated by removing the generated cell from the boundary and adding its neighbors. The process stops when all the cells are visited and the actual boundary of the profile is reached. The cellular multiplication with no member removal results in a structure with 95 self-stress states among which 70 states correspond to constitutive cells and 25 to virtual cells.

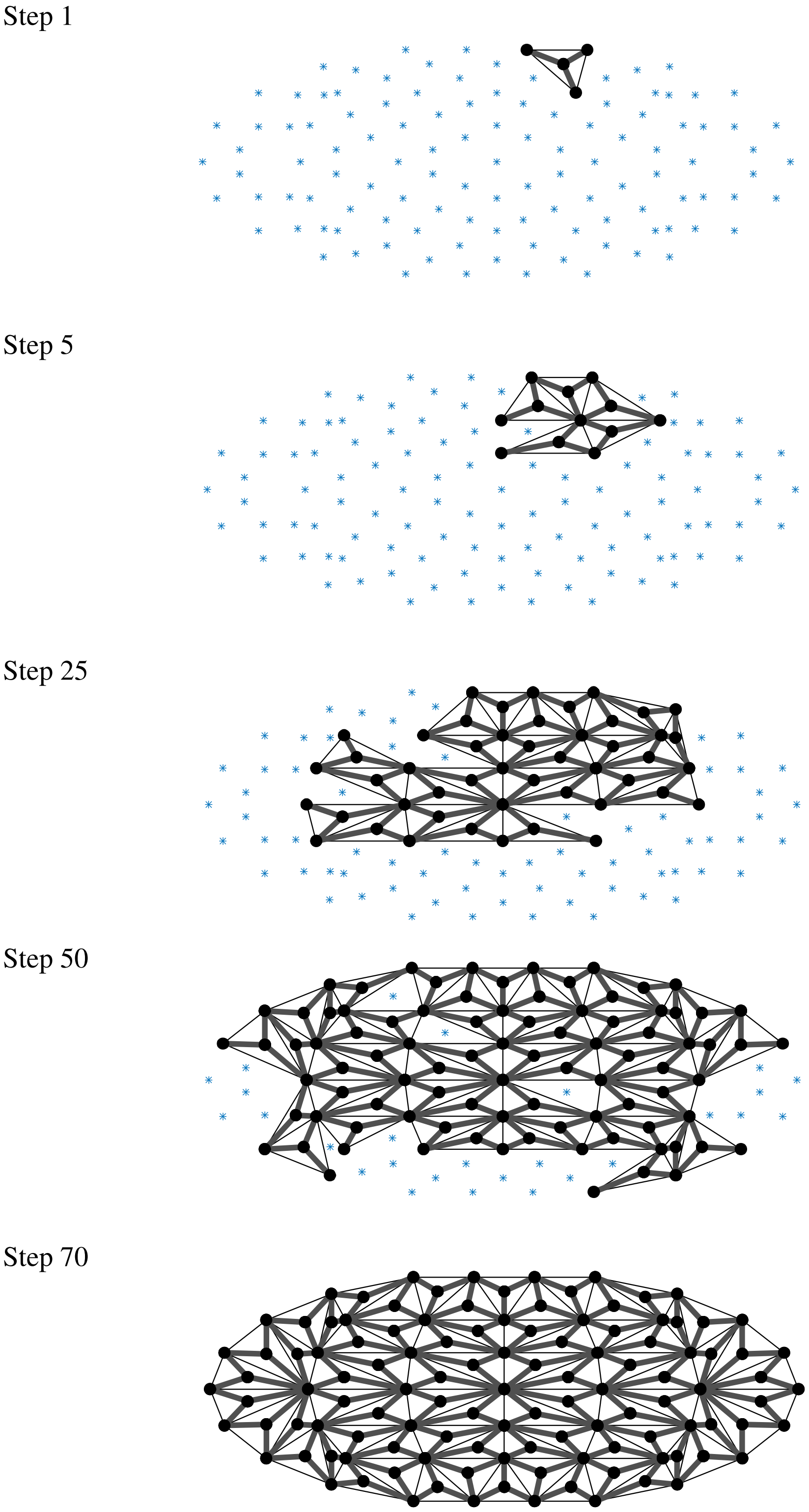

Figure 17: Snapshots during the generation process of a large elliptical tensegrity structure.

## 7. Discussion

Cellular multiplication reflects a generative design scheme for large tensegrity structures based on their composition from stable sub-structures. Through the combination of topology definition and form finding, the proposed scheme provides control over the nature as well as the number of self-stress states in a planar tensegrity structure. It also enables the exploration of large irregular tensegrity systems. Moreover, the form of the self-stress matrix resulting from the proposed method and its direct association to the minimally rigid tensegrity units provides designers with additional information on the self-stress space describing the stability of structure, that cannot be obtained through the analysis of the rank of its equilibrium matrix.

The method is based on gluing operations of rigid graphs along at least $d$ nodes, where $d$ is the dimension of the workspace. According to Whitely [38], gluing operations in this case preserves the rigidity of the resulting graph. When the set of shared nodes between two cells is less than the dimension $d$ of the workspace, the resulting structures will include mechanisms such as the rotation illustrated in Figure18.

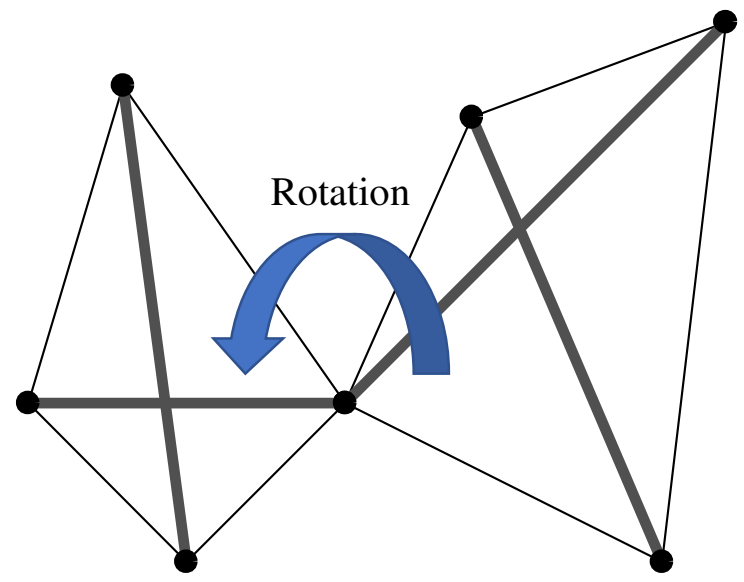

Figure 18: mechanism creation when number of shared nodes is 1 in the plane

Maxwell rule [48] gives that if $s$ is the number of self-stress states in the structure, $m$ is the number of internal infinitesimal mechanisms and $B$ is the Laman bound of the graph of the structure then:

$$s - m = B \tag{25}$$

Infinitesimal mechanisms have to be accounted for in such cases and taken into consideration in the construction of a basis for the self-stress space. As an example, consider the structure in Figure 19. The structure on the left is initially composed of 8 cells attached through one node. The structure has thus 5 non-trivial mechanisms. The addition of the central cell stabilizes the structure removing all 5 mechanisms, while also creating an additional state. If proposition III was employed without taking into consideration the existence of infinitesimal mechanisms in the structure, the addition of the central cell would supposedly add 6 self-stress states, one corresponding to the real cell and 5 to virtual cells. However, in this case the existence of mechanisms eliminates the self-stress states from the virtual cells.

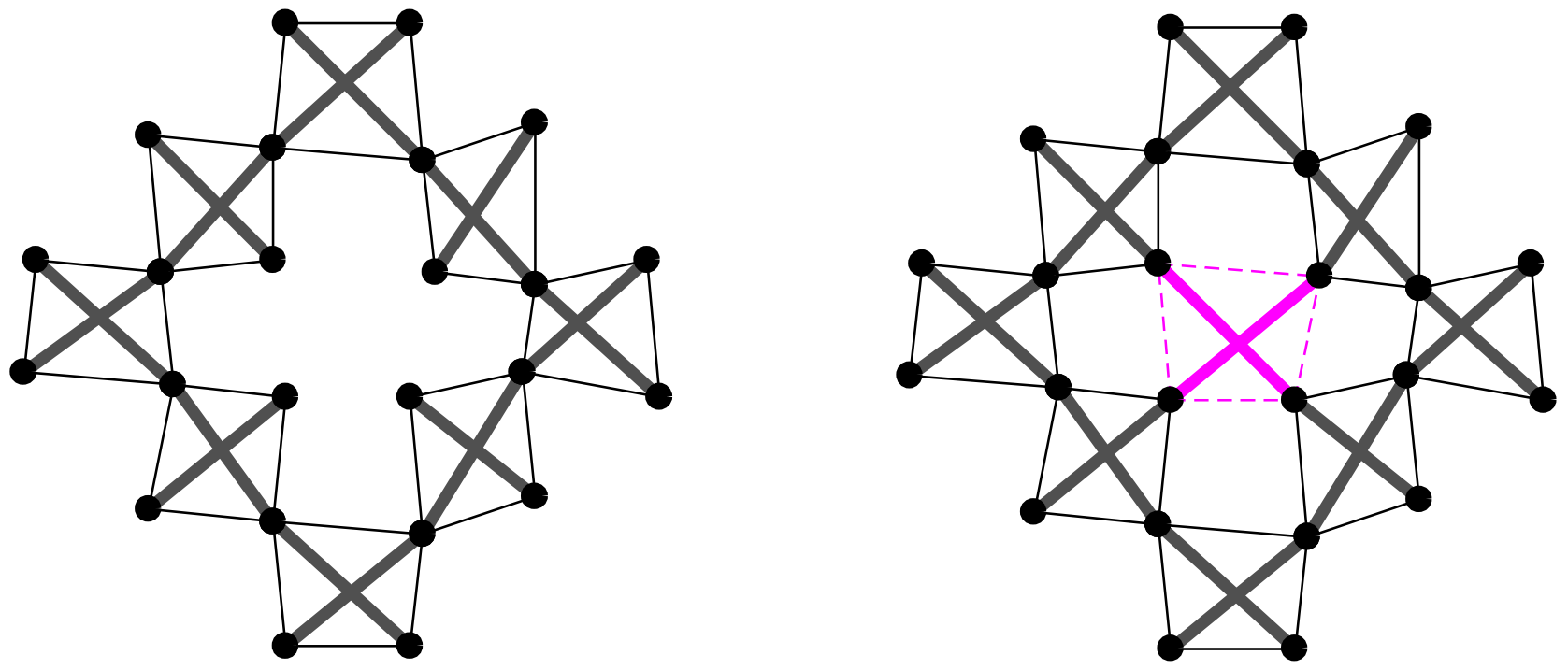

Figure 19: Cellular multiplication allowing mechanisms.

It should also be noted that the method can be extended to include the application of external loads and for the generation of three-dimensional tensegrity systems. The application of external loads will result in a two-component stress expression with the first component being a homogeneous solution reflecting the self-stress (given by the current formulation), and a particular solution reflecting the influence of the applied load on the system. For the generation of three-dimensional tensegrity systems, the complete graphs $K_5$ on five nodes should be considered which increases the complexity of the method.

## 8. Conclusions

In this paper, a novel method for the generation of planar tensegrity structures is presented. The method is based on tensegrity cells (infinitesimally rigid structures with one self-stress state that can compose any tensegrity structure) and is inspired by the multiplication mechanisms of unicellular organisms. It is shown that changes in the dimension of the self-stress space of a cell-composed structure between generation steps depend on the number of nodes being added, as well as the number of shared and removed members with additional self-stress states originating from newly generated cells and their interaction with existing cells. Cell interaction is explained through the concept of virtual cells: stable sub-structures with a single self-stress state composed of elements from the different cells but without using all the elements of a given cell. A method for identifying virtual cells and composing a basis of the self-stress space is also proposed. Cellular multiplication enables the control of key design parameters in tensegrity structures such as the equilibrium shape, number of nodes and members, as well as the self-stress. It can thus serve as a powerful and flexible tool for the generative design of planar tensegrity structures.

## 9. Acknowledgements

This material is based upon work supported by the National Science Foundation under grant no. 1638336. David Orden has been partially supported by MINECO Projects MTM2014-54207 and MTM2017-83750-P, as well as by H2020-MSCA-RISE project 734922 - CONNECT.